\pdfoutput=1

\documentclass[%
    reprint,
    showpacs,preprintnumbers,
    showkeys,
    amsmath,amssymb,
    aps,
    prx,
    floatfix,
    longbibliography,
    notitlepage,
]{revtex4-2}

\usepackage[english]{babel}
\usepackage[utf8]{inputenc}
\usepackage{graphicx}
\usepackage{bm}
\usepackage{hyperref}
\usepackage{xcolor}
\usepackage{epsfig}
\usepackage{amsmath}
\usepackage{amsfonts}
\usepackage{amssymb}
\usepackage{amsthm}
\usepackage{gensymb}
\usepackage{enumerate}
\usepackage{tikz}
\usepackage{dsfont}
\usepackage[all]{nowidow}
\usepackage{pifont}
\usepackage{algorithm}
\usepackage{algpseudocode}
\usepackage[capitalise]{cleveref}
\usepackage[export]{adjustbox}
\usepackage{xspace}
\usepackage{nicefrac}

\usepackage{doi}

\usepackage[letterspace=130]{microtype}

\definecolor{refColor}{HTML}{0376E9}
\definecolor{figColor}{HTML}{0376E9}
\definecolor{urlColor}{HTML}{0376E9}

\hypersetup{%
    unicode=false,                 % non-Latin characters in Acrobat’s bookmarks
    pdftoolbar=true,               % show Acrobat’s toolbar?
    pdfmenubar=true,               % show Acrobat’s menu?
    pdffitwindow=false,            % window fit to page when opened
    pdfstartview={FitH},           % fits the width of the page to the window
    pdfpagemode=false,             %toggle sidebar after compiling
    pdftitle={},                   % title
    pdfauthor={Simon Stastny},      % author
    pdfsubject={Quantum physics},  % subject of the document
    pdfcreator={Nicolai Lang},     % creator of the document
    pdfproducer={Simon Stastny},    % producer of the document
    pdfkeywords={},                % list of keywords
    pdfnewwindow=true,             % links in new window
    colorlinks=true,               % false: boxed links; true: colored links
    linkcolor=figColor,            % color of internal links
    citecolor=refColor,            % color of links to bibliography
    filecolor=magenta,             % color of file links
    urlcolor=urlColor              % color of external links
}

\newcommand{\bra}[1]{\mathinner{\langle{#1}|}}
\newcommand{\ket}[1]{\mathinner{|{#1}\rangle}}
\newcommand{\braket}[1]{\mathinner{\langle{#1}\rangle}}

\renewcommand{\vec}[1]{\mathbf{#1}}

\renewcommand{\vec}[1]{\boldsymbol{#1}}

\newcommand{\sz}{\sigma_z}

\newcommand{\sx}{\sigma_x}
\newcommand{\tz}{\tau_z}

\newcommand{\tx}{\tau_x}

\tikzstyle{roundlabel}=[circle,inner sep=1.5pt,fill=tikzgrey,color=black,text=white,draw,font=\sffamily\footnotesize]
\tikzstyle{TlabelB}=[rectangle callout,rounded corners=0.03cm,inner sep=3.0pt,fill=tikzlight,color=tikzlight,text=tikzgrey,draw,font=\sffamily\footnotesize,callout absolute pointer={#1},at={#1},above=0.2cm,align=center]
\tikzstyle{smallbullet}=[circle,inner sep=0.5pt,fill=tikzorange,color=tikzorange,text=white,draw,font=\sffamily\small\bfseries]
\tikzstyle{labelline}=[rounded corners=0.5mm,black,line width=2pt,solid]
\tikzstyle{labeltext}=[color=white,fill=white,inner sep=0.5pt,text=tikzgrey,draw,font=\sffamily\footnotesize]
\tikzstyle{math}=[font=\Large,align=center]
\tikzstyle{figure}=[anchor=center]
\tikzstyle{label}=[circle,inner sep=0.07cm,fill=white,text=black,draw=white]
\tikzstyle{tlabel}=[inner sep=0.07cm,text=black]
\tikzstyle{wlabel}=[inner sep=0.07cm,text=black,anchor=west,font=\footnotesize]
\tikzstyle{figtext}=[anchor=west, rounded corners=0.0cm,inner sep=0.07cm,fill=white,color=white,text=black,draw,font=\sffamily\footnotesize]

\definecolor{sublatticecolor}{HTML}{CCCCCC}

\definecolor{optimized}{HTML}{e4000d}
\definecolor{unoptimized}{HTML}{000000}

\graphicspath{{./fig}{./data}}

\begin{document}

\title{The singlet-triplet and exchange-only flopping-mode spin qubits}

\author{Simon Stastny}
%\email{simon.stastny@uni-konstanz.de}
\author{Guido Burkard}
%\email{guido.burkard@uni-konstanz.de}
\affiliation{%
    Department of Physics, University of Konstanz, D-78457, Konstanz, Germany
}

%\date{\today}

%+++++++++++++++++++++++++++++++++++++++++++++++++++++++++++++++++++++++++++++++
%% ABSTRACT
%+++++++++++++++++++++++++++++++++++++++++++++++++++++++++++++++++++++++++++++++

\begin{abstract}
	Semiconductor-based spin qubits embedded into a superconducting microwave
	cavity constitute a fast-progressing and promising platform 
    for realizing fast and fault-tolerant qubit control with long-range two-qubit coupling. 
    The flopping-mode spin
	qubit consists of a single electron in a double quantum dot; it combines a charge qubit with a spin qubit. With its strong and tunable cavity coupling, the flopping-mode qubit is proven to be well-suited for low-power qubit control and
	cavity-mediated long-range quantum gates. 
    The singlet-triplet (ST) and exchange-only (EO) qubits are multi-electron realizations that go without broadband control and are protected from some types of noise, but are challenging to couple to each other and to microwave cavities.
    We combine the flopping-mode concept with the ST and EO qubits and propose two new
	  flopping-mode qubits that 
	consist of three (four) quantum dots, 
	occupied by two (three) electrons near the 
	$ (1,0,1) \leftrightarrow (0,1,1) $
    [$ (1,0,1,1) \leftrightarrow (0,1,1,1) $]
	charge transition. The two-electron system augments the $ST_0$ spin qubit with a charge qubit
	that interacts transversally and longitudinally with a cavity. 
    Both couplings are highly tunable, and the longitudinal coupling 
    distinguishes the flopping-mode ST qubit from the regular flopping-mode qubit. 
    The longitudinal coupling allows for non-dissipative universal
	control similar to superconducting transmon qubits. 
    The EO flopping-mode qubit comprises four dots 
	occupied by three electrons and opens a new possibility to perform two-qubit gates for EO qubits that are challenging to perform directly with the exchange coupling. 
    We use input-output theory to provide means of extracting the coupling strengths
    from cavity transmission data.

\end{abstract}

\maketitle

%+++++++++++++++++++++++++++++++++++++++++++++++++++++++++++++++++++++++++++++++
\section{Introduction}
%+++++++++++++++++++++++++++++++++++++++++++++++++++++++++++++++++++++++++++++++

Spin qubits in semiconductor quantum dots (QDs)  \cite{LossDiVincenzoPhysRevA.57.120}
are a promising platform for
quantum computation and have called a flourishing research field into action.
The absence of a nuclear spin in
${}^{28}$Si allows for qubits with 
remarkably long coherence times~\cite{Veldhorst_coherencetimeslong_veldhorst2014addressable,Coherencetimeslong2_tyryshkin2012electron} and quantum gates with error rates in the range or below typical error thresholds for fault-tolerant quantum computation \cite{wu2024hamiltonian}.

Various qubit encodings beyond the single-spin (Loss-DiVincenzo) qubit
have been realized within this platform,
such as the singlet-triplet ($ST_0$) qubit~\cite{singlet_triplet_Kane1998_kane1998silicon,Levy2002_PhysRevLett.89.147902,Petta2005_doi:10.1126/science.1116955}
and the exchange-only (EO) qubit~\cite{divincenzo2000universal,FongWandzureEOquantumcomputingDFSubspace,Baconetal2000_PhysRevLett.85.1758,taylor2005fault,Taylor_srinivasa_PhysRevLett.111.050502,Medford_PhysRevLett.111.050501}. These encodings require more than one quantum dot for each qubit, and in return
allow for partial or complete electric baseband control, thus relaxing the requirements regarding high-frequency driving or inhomogeneous magnetic fields.
%
%have some advantages over single-spin qubits, 
%as these qubits are
%not localized in one dot, making them more %resistant against
%local magnetic noise~\todo{citations?} 
%
In addition, the EO qubit
allows for complete control using only the 
exchange interaction~\cite{divincenzo2000universal,weinstein2023universal}.
However, this simplification comes with a substantial complexity of the implementation of exchange-based two-qubit gates \cite{divincenzo2000universal,FongWandzureEOquantumcomputingDFSubspace,ivanova2024discovery}.

Universal quantum computing requires non-local two-qubit gates, in addition to the nearest-neighbor gates that can be directly implemented with the exchange interaction. 
Multi-spin exchange-coupled quantum dot
systems can also be used to realize a spin bus, a
spin chain which can be used as
an intermediate-range two-qubit coupling\cite{BoseSpinBus_PhysRevLett.91.207901,bose2007quantum,friesen2007efficientspinbus,sigillitoPetta2019coherent_bus}.
Moreover, super-exchange can be used to couple two spin qubits
by placing them next to a mediator unit, (virtual) 
excitations of which can mediate an indirect exchange interaction
between the spins of interest~\cite{craig2004tunable,malinowski2019fast}. 
Long-range interactions between EO qubits
are hard to realize, as the exchange interaction only couples
nearest neighbor spins \cite{SpinQubit_Review_RevModPhys.95.025003}.
Spin shuttling is an effective way of realizing non-local
spin-spin interactions by transporting spins with a modulating
confinement field \cite{baart2016single,fujita2017coherent}.
 
Much longer-range interactions between spin qubits may be achievable by leveraging the interactions between localized spins and delocalized cavity photons in a superconducting microwave cavity \cite{burkard2020superconductor_EDyn}.
To achieve sufficiently large spin-photon couplings, substantial electric dipole couplings are required. 
Although the two-electron ST qubit and the three-electron EO qubit possess an electric dipole near a (1,1)-(0,2) or (1,1,1)-(1,0,2) charge transition \cite{Longrange1childress2004mesoscopic,Longrange2burkard2006ultra,UngererbottcherSTlong2022parametric,russ2016coupling,landig2018coherent}, the corresponding dipoles are relatively small.
However, by adding an empty dot, charge-like degrees of freedom (charge qubits) can be integrated into spin-qubit systems. 
A paradigmatic example of such a spin-charge qubit system is the flopping-mode spin qubit \cite{BenitoFloppingmode2_NoiseTheoExpPhysRevB.100.125430,CrootPhysRevResearch.2.012006} which extends the Loss-DiVincenzo (LD) qubit to one electron in two dots, and lends itself to the
realization of fast long-range
interactions without charge transport. Due to the exchange of virtual photons,
distant two-qubit operations can be realized \cite{longbenito2019optimized,warren2019long}. 
The coupling can be achieved because
the charge-like qubit can couple to the electric field of
a superconducting cavity, and establish coherent photon charge
interactions. Then, due to an artificial spin-orbit interaction, photon-spin interactions can arise
 \cite{CavityQEDmi2017strong,hu2012strong,mi2018coherent,PhysRevB.96.235434,longrange2borjans2020resonant,harvey2022coherent},
 allowing for non-local two-qubit gates \cite{dijkema2024cavity}.

By injecting a probe field into the cavity and measuring the output field, 
cavity QED can also be used to measure these couplings, or 
to perform a dispersive readout of the qubit 
states \cite{d2019optimal,petersson2010charge,mielke2021nuclear,house2015radio,colless2013dispersive}. In the case of the 
flopping-mode spin qubit \cite{PhysRevB.96.235434,GE_Floppingmode_Exp_yu2023strong,Petta_Jonas_entagledFloppingmodesPhysRevB.110.035304,FloppingmodeinGE_Mutter_-PhysRevResearch.3.013194} the experimental realization was successful \cite{FloppingExpChina10.1063/5.0137259},
and entanglement between two cavity-coupled 
qubits has been measured. \cite{Petta_Jonas_entagledFloppingmodesPhysRevB.110.035304}.
Electrically tunable 
tunneling and detuning of the dots give the
system a dipole moment and a magnetic
field gradient between the dots gives rise to the 
spin-orbit coupling. This flopping-mode qubit thus
is a mechanism of making spin qubits like the LD qubit
`charge like'. The question arises as to whether this is 
possible for more general spin qubits such as the more noise 
resistant
delocalized multi electron $ST_0$- or EO-spin qubits, i.e., whether 
the ST and EO qubits can be made `charge like' in this way,
by coupling them to an additional unoccupied quantum dot. 
In this paper, we introduce two new spin-qubit flavors that combine the flopping-mode qubit with the $ST_0$ and the EO qubits and inherit some of the favorable properties of both concepts.

This is done by equipping the singlet-triplet qubit
and the EO qubit with an additional, unoccupied quantum dot. Starting from the $ST_0$ qubit,
the resulting qubit then consists of three quantum dots in a magnetic field gradient. 
The system operates near the $(0,1,1) \leftrightarrow (1,0,1)$ 
charge transition, endowing it with an electric dipole moment. Such an arrangement can also be used for a cavity-based
measurement of the exchange interaction in adjacent quantum dots~\cite{PhysRevResearch.4.033048}.
The flopping-mode EO qubit comprises four quantum dots
in a global Zeeman field. Charge hybridization between the 
$(0,1,1,1), ~(1,0,1,1)$, and $(1,1,0,1)$ charge states provides the qubit with an electric dipole moment.
With this approach, one can 
use the advantages of 
these two qubit types, while
achieving a tunable transversal and longitudinal 
cavity coupling. Especially longitudinal coupling is
a promising feature as it is not subject to the Purcell effect
and has already been investigated both in superconducting systems such as 
transmon qubits~\cite{blais2021circuit,didier2015fast,jin2012strong,transmonricher2017inductively}
and in spin qubits embedded in superconducting cavities \cite{ruskov2019quantum,ruskov2021modulated}. Moreover, as we will show, the longitudinal coupling enables a coupling of the EO subsystem (rather than subspace) qubit to cavity photons. 

Recently, a similar system, consisting of a triple quantum dot occupied with two electrons
has been studied in ~\cite{TDQresonatorsehrahnlich_PhysRevResearch.6.043029,friesen2007efficientspinbus}. These works 
investigate 
the use of the system as two qubits, a single LD qubit 
and a flopping-mode qubit, that can couple locally, 
while the flopping-mode 
qubit is coupled to a resonator. 

The remainder of this paper is structured as follows: In \cref{sec:model}
we introduce the model and we investigate and derive an
effective qubit Hamiltonian for the flopping-mode ST qubit. 
In \cref{sec:resultseffst} we analyze the spin-photon couplings from the effective
Hamiltonian. In \cref{sec:inputoutput} we derive an equation for the cavity
transmission and phase and propose an experiment
to obtain the spin-photon couplings by analyzing
these two quantities.
We consider the flopping-mode EO qubit in \cref{sec:floppingexchange}. 
Finally, we summarize our results in \cref{sec:summary}.

%+++++++++++++++++++++++++++++++++++++++++++++++++++++++++++++++++
\section{Model of the singlet-triplet flopping-mode qubit}
\label{sec:model}
%+++++++++++++++++++++++++++++++++++++++++++++++++++++++++++++++++

%///////////////////////////////////////////////////////////////////////
\begin{figure}[tb]
	\centering
	\includegraphics[width=0.7\linewidth]{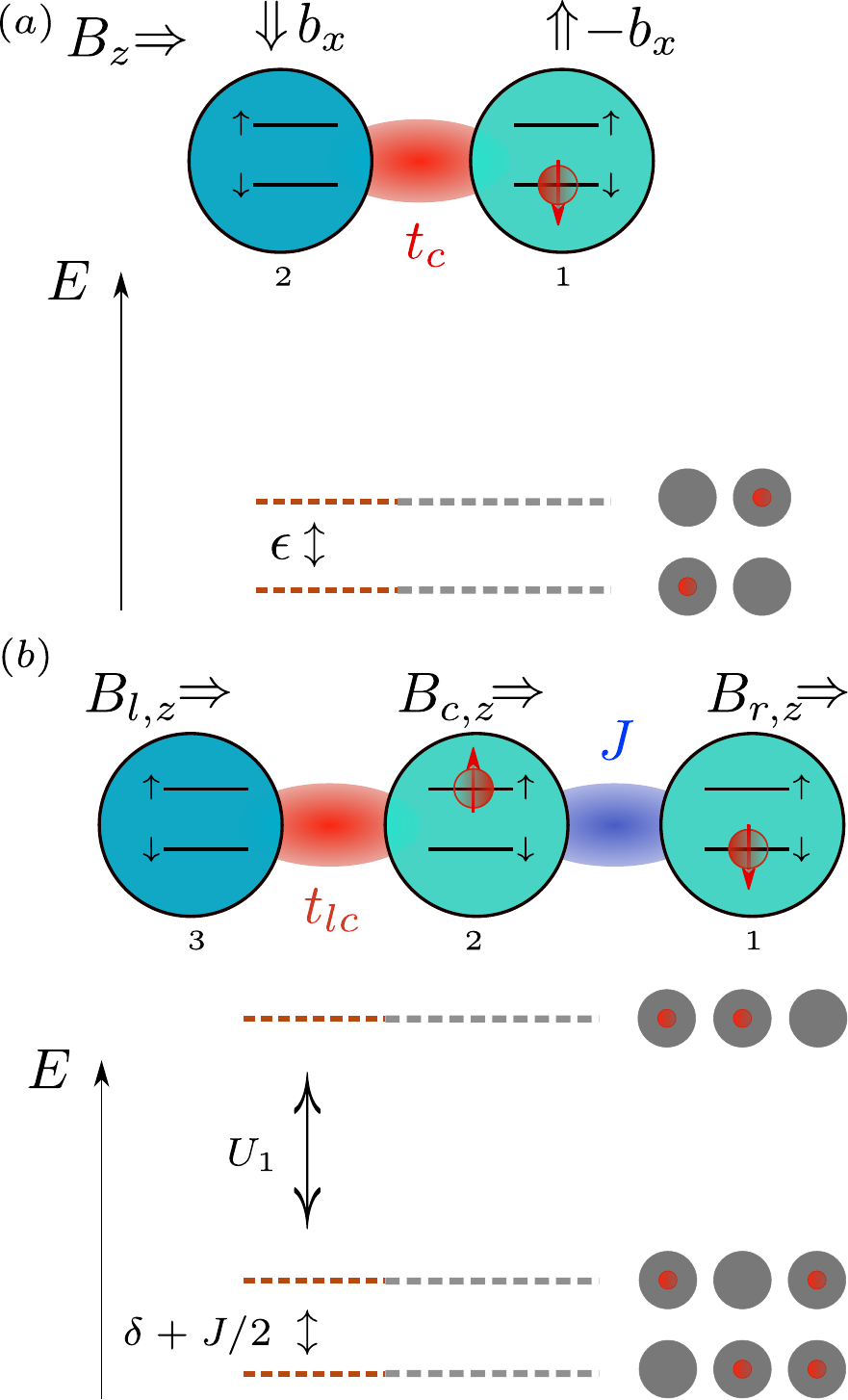}
	\caption{%
	\emph{Flopping-mode spin qubits.}
	(a) Schematics and charge energy levels of the (single-spin) flopping-mode qubit. 
    A double quantum
    dot (DQD) occupied by one
    electron is placed in a global Zeeman field $B_z$ in the 
    chosen $z$-direction. A perpendicular magnetic field gradient
    $b_x$ in $x$-direction is applied. 
    The tunnel coupling $t_c$ and the field gradient induce an effective (synthetic) spin-orbit (SO) coupling.
    An energy splitting of the two charge states can be
    achieved by detuning the two dots by $\epsilon$. If $|\epsilon|\gg |t_c|$, 
    the electron is confined in one dot, effectively acting as an LD qubit.
    If $|\epsilon|\ll |t_c|$, the electron is delocalized, thus
    allowing for SO-coupling (flopping mode).
    (b) The proposed singlet-triplet flopping-mode qubit. A third dot (3) is attached to a 
    singlet-triplet qubit (dots 1 and 2). Local magnetic fields
    $B_{i,z}$ are applied to the three dots and form a field gradient.
    One of the three charge states is 
    separated by the nearest-neighbor Coulomb repulsion energy $U_1$, the other two again form an effective
    flopping-mode qubit, detuned by $\delta+J/2 \ll U_1$.
    }
	\label{fig:thesetup}
\end{figure}
\begin{figure}[tb]
	\centering
	\includegraphics[width=0.8\linewidth]{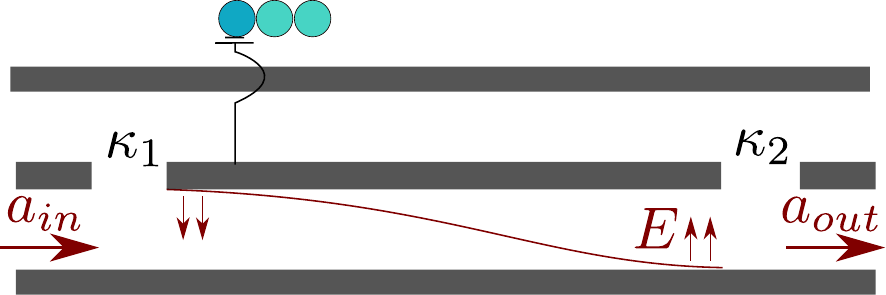}
	\caption{%
	\emph{Schematics of the cavity setup.}
    The cavity can leak at the two exit ports 1 and 2 with leakage rates
    $\kappa_1$ and $\kappa_2$.
    The third quantum dot of the singlet-triplet flopping-mode qubit is coupled to the voltage at the center conductor of the cavity at an antinode of the electric field $E$ (near port 1). The cavity is probed by injecting a microwave tone $a_{\rm in}$ and observing the transmitted signal $a_{\rm out}$. The electric-field profile $E$ is indicated in red.}
	\label{fig:thesetupcavityst0}
\end{figure}
%///////////////////////////////////////////////////////////////////////
%
The starting point of our model is the 
flopping-mode qubit \cite{PhysRevB.96.235434},
consisting of one electron in a double quantum dot, see \cref{fig:thesetup}(a),
with both dots in a global Zeeman field $B_z$. 
The electron can tunnel between the two ground state orbitals
of the dots, with tunnel matrix element $t_c$. A
perpendicular local magnetic field gradient $\Delta b= 2 b_x$ is applied
to realize spin-orbit hybridization. The ground state energies of the dots differ by an energy $\epsilon$. With the electron spin $\bm S = \frac{1}{2}\bm \sigma$ and the charge degree of freedom
$\bm \tau$, the Hamiltonian for this model is
given by 
\begin{align}
	H_{{\rm fm}, 1} = \frac{\epsilon}{2}\tau_z+t_c \tau_x + \frac{B}{2}\sigma_z+
	\frac{\Delta b}{2}\sigma_x\tau_z+g_c\tau_z(a+a^\dagger),
\end{align}
where $\tau_x$ and $\tau_z$ are Pauli matrices defined in the $\ket{L,R}$ basis,
while $\sigma_x$ and $\sigma_z$ are Pauli matrices for the electron spin. 
The last term describes the coupling of the charge to a 
microwave cavity mode with coupling strength $g_c$. 
The charge parameters $\epsilon$ and $t_c$ are electrically
tunable and determine the effective
spin-photon coupling~\cite{PhysRevB.96.235434}. The maximum $g_c$ is reached at the charge
sweet spot where $\epsilon=0$. The flopping-mode qubit can be seen as an LD qubit~\cite{LossDiVincenzoPhysRevA.57.120} extended by an
additional tunnel-coupled empty dot that enables
spin-photon coupling. We extend this idea of the flopping-mode qubit to the singlet-triplet $ST_0$ qubit \cite{Petta2005_doi:10.1126/science.1116955} by adding an empty dot. 
The $ST_0$ qubit is realized by two exchange-coupled quantum dots occupied by
one electron each, immersed into an inhomogeneous magnetic (Zeeman) field.
To apply the `flopping-mode mechanism', i.e., endow 
this qubit with an electric dipole moment, 
this $ST_0$ qubit is coupled to a third, 
empty quantum dot.
The resulting model consists of three quantum dots
occupied by a total of two electrons, see \cref{fig:thesetup}(b)
. 
The local Zeeman field at each dot $j\in \{l,c,r\}$ is denoted as $B_{z,j}$. 
The onsite energies of each dot and 
tunnel couplings $t_{lc}$ and $t_{cr}$ between dots
can be controlled electrically. 
The Hamiltonian is given by
\begin{equation}
\begin{split}
H_{{\rm fm}, 2} = & \biggl[\frac{J}{4}\biggl(\vec\sigma_1\cdot \vec\sigma_2-1\biggr)+\frac{1}{2}\vec B_c\cdot\vec \sigma_1\biggr]\frac{1-\tau_z}{2}\\
& +\biggl[\delta + \frac{1}{2}\vec B_l\cdot \vec \sigma_1\biggr]\frac{1+\tau_z}{2}+t_{lc}\tau_x+\frac{1}{2}\vec B_r\cdot\vec\sigma_2 ,
\label{eq:singlet-triplethamkomplett}
\end{split}
\end{equation}
where $\vec S_i=\frac{1}{2}\vec \sigma_i$
are the spins of the two electrons
($\hbar =1$, $i=1,2$) and $J$ is the 
exchange interaction between dots 1 and 2, see \cref{fig:thesetup}(b).
The onsite energies can be chosen, such that the charge 
configurations $(1,0,1)$ and $(0,1,1)$ are detuned by a controllable 
detuning $\delta \ll U_1$, where $U_1$ is the next-neighbor Coulomb repulsion. A third charge configuration
$(1,1,0)$, as well as charge configurations comprising double occupation of dots, are energetically separated by either $U_1$ or the on-site Coulomb energy, and are thus neglected.
The system operates near the $(0,1,1)\leftrightarrow (1,0,1)$ 
charge transition.
Away from the charge transition, i.e., for
$\delta \gg |t_{lc}|$, the system is in 
the $(0,1,1)$ charge state and the electron spins couple 
with the exchange interaction $J$. 
This charge state corresponds to a $ST_0$ qubit
next to an empty dot. 
Inside the charge transition regime, i.e. for $\lvert\delta\rvert \ll |t_{lc}|$, the charge states hybridize.
In the singlet-triplet basis of the two-spin subspace, the Hamiltonian \cref{eq:singlet-triplethamkomplett} restricted to the computational 
$ST_0$ subspace is found to be
\begin{equation}
    \begin{split}
H_\text{sys}=&\frac{\delta+J/2}{2}\tz + t_{lc} \tx 
+\frac{B_{l,z}-B_{c,z}}{4}\tilde\sigma_x\tz\\
&+\frac{B_{l,z}+B_{c,z}-2B_{r,z}}{4}\tilde\sigma_x 
+ \frac{J}{2}\tilde \sigma_z\frac{1-\tau_z}{2},
        \label{eq:Htqdsinglettriplet}
    \end{split}
\end{equation}
where the spin operators $\tilde \sigma_{z}$ and 
$\tilde \sigma _x$ 
act on the singlet and triplet states $\ket{S}=(\ket{\uparrow\downarrow}-\ket{\downarrow\uparrow})/\sqrt{2}$ and $\ket{T_0}=(\ket{\uparrow\downarrow}+\ket{\downarrow\uparrow})/\sqrt{2}$ as
$\tilde\sigma_z\ket{S}=-\ket{S}$ and $\tilde\sigma_z\ket{T_0}=\ket{T_0}$.
In the following, we define the gradient fields
\begin{gather}
    \frac{\Delta b}{2} = \frac{B_{l,z}-B_{c,z}}{4},\\
    \frac{\Delta B}{2}=\frac{B_{l,z}+B_{c,z}-2B_{r,z}}{4}.
\end{gather}
Analogously to the flopping-mode qubit, the two low-energy charge
states are coupled to the electric field of the microwave cavity mode 
$H_c = \omega_c a^\dagger a$ via the electric-dipole Hamiltonian
\begin{equation}
    \begin{gathered}
    H_I=g_c \tau_z\left(a+a^\dagger\right),
        \label{eq:Hi}
    \end{gathered}
\end{equation}
where $g_c$ is the charge-photon coupling. In the
following, we set $g_c=0.5~\mu \rm{eV}=120\,{\rm MHz}$. The 
cavity-qubit system is depicted in \cref{fig:thesetupcavityst0}.
The cavity has two ports, port 1 and port 2, 
that are described by their photon-loss rates $\kappa_1$
and $\kappa_2$, and the total decay rate is defined as $\kappa=\kappa_1+\kappa_2+\kappa_{\rm int}$ with $\kappa_{\rm int}$ the internal photon loss.
A probe field \cite{PhysRevResearch.4.033048} with frequency 
$\omega_r$ defined with the Hamiltonian
\begin{equation}
    \begin{gathered}
    H_p= i\sqrt{\kappa_1}\left(a_{\rm in}e^{-i\omega_r t}a^\dagger+
    a_{\rm in}^*e^{i\omega_r t}a\right)
        \label{eq:Hp}
    \end{gathered}
\end{equation}
can be applied at port 1 (which we define to be the input port).
In summary, the entire cavity-qubit system is thus described by
the Hamiltonian
\begin{equation}
    \begin{gathered}
    H=H_\text{sys}+H_c+H_I+H_p.
        \label{eq:Hsys}
    \end{gathered}
\end{equation}
%

%+++++++++++++++++++++++++++++++++++++++++++++++++++++++++++++++++
\section{Effective Hamiltonian}
\label{sec:resultseffst}
%+++++++++++++++++++++++++++++++++++++++++++++++++++++++++++++++++

\begin{figure}[tb]
	\centering
	\includegraphics[width=0.8\linewidth]{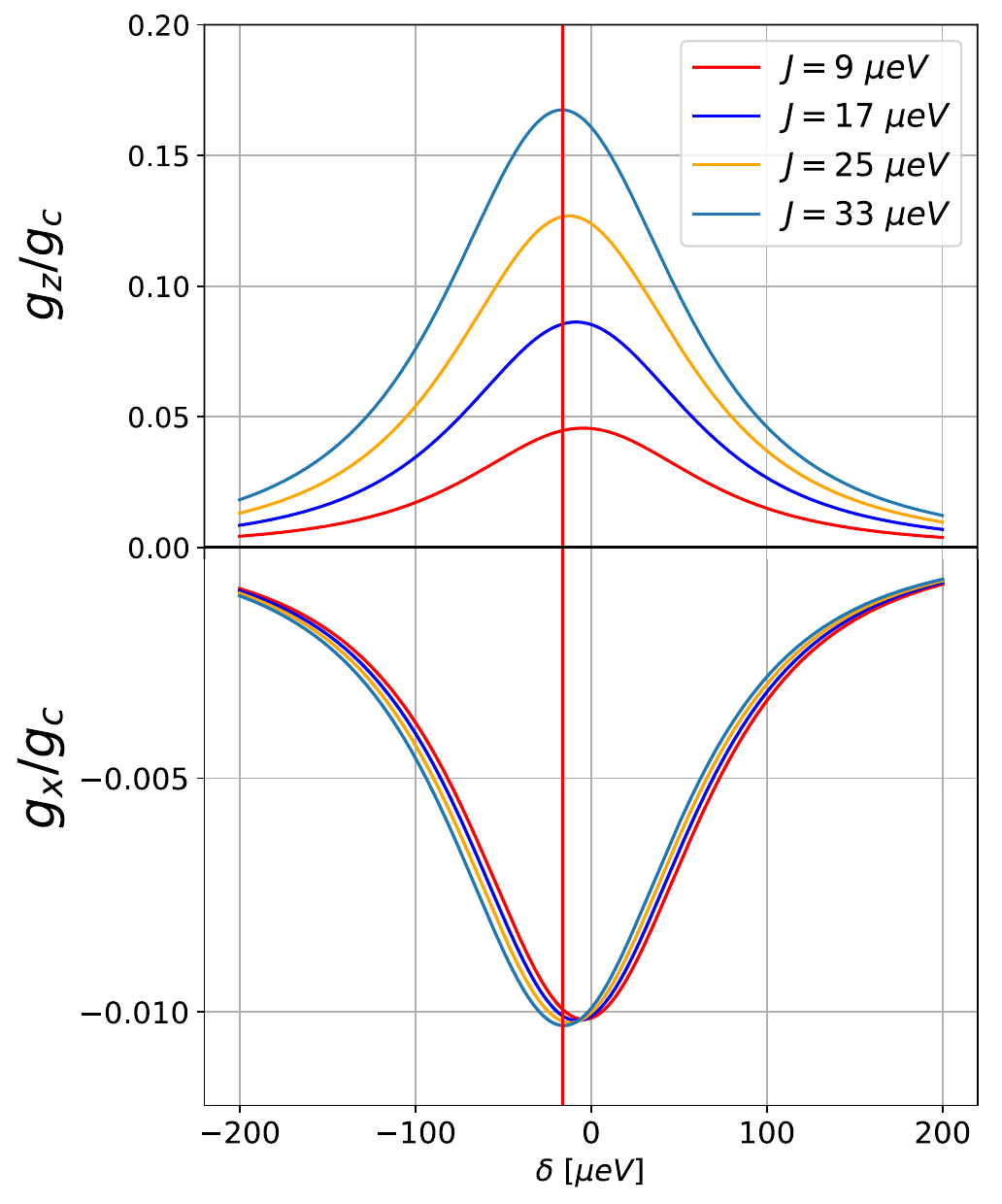}
	\caption{%
	\emph{Cavity couplings $g_z$ and $g_x$ for the $ST_0$ flopping-mode qubit.}
    The effective spin-photon
    couplings are plotted in units of $g_c$. Note the opposite sign of $g_z$ and $g_x$.
    The coupling strengths are maximal if the qubit resonance is near $-\delta=J/2$, and decays as $\delta$ increases.
    At $\delta = -16.527\,\mu{\rm eV}$ (vertical red line) we obtain $g_z/g_c=0.168$ and 
    $g_x/g_c=-0.010$.
    Parameters used for the plot are $t_{lc}=50~\mu {\rm eV}$, $J=33~\mu {\rm eV}$,
    $B_{l,z}=7~\mu {\rm eV}$, $B_{c,z}=5~\mu {\rm eV}$, and $B_{r,z}=2~\mu {\rm eV}$.
    }
	\label{fig:cavitycouplings}
\end{figure}

We diagonalize
the charge part $(\delta+J/2)\tau_z/2 + t_{lc}\tau_x$ of the Hamiltonian \cref{eq:Htqdsinglettriplet} to obtain the charge 
eigenenergies $\Omega=\pm \sqrt{(\delta+J/2)^2+4t_{lc}^2}/2$. 
Next, a Schrieffer-Wolff transformation 
is applied to $H$,
projecting on the low-energy charge sector. 
The resulting perturbation theory is valid in the regime
$g_c \ll \Delta b,J \ll \Omega$ where 
terms of order  $O(g_c^2/\Omega)$ can be neglected.
Note that $H_p$ is invariant under that transformation.
The resulting effective Hamiltonian is given by
\begin{equation}
    \begin{split}
H_\text{eff}=&\left[\frac{J_\text{eff}}{2}+g_z(a+a^\dagger)\right]\hat{\sigma}_z\\
&+\left[\frac{b_\text{eff}}{2}+g_x(a+a^\dagger)\right]\hat{\sigma}_x
+\omega_c a^\dagger a +H_p.
        \label{eq:Heffst0}
    \end{split}
\end{equation}
The dispersive shift and corrections due to vacuum 
fluctuations are of order $g_c^2$ and are thus neglected here. 
The Pauli matrices $\hat{\sigma}_{x}$ and $\hat{\sigma}_z$ act on the 
dressed spin states. 
The terms $J_\text{eff}$ and $b_\text{eff}$ are the effective qubit parameters which are given by
\begin{equation}
    \begin{split}
J_\text{eff}=&\frac{J}{2}(1+\sin\theta)\\
& + \frac{J^3}{4}\frac{\cos^2\theta\sin\theta}{J^2\sin^2\theta-4\Omega^2}
 + J\Delta b ^2\frac{\cos^2\theta}{J^2-4\Omega^2},\\
%&+J\cos^2\theta\left(\frac{(J^2/8) \sin\theta(J^2-4\Omega^2)
%+\frac{(\Delta b)^2}{2}(J^2\sin^2\theta-4\Omega^2)}
%{(J^2-4\Omega^2)(J^2\sin^2\theta-4\Omega^2)}\right),\\
b_\text{eff}=&-\Delta b_{lc}\sin\theta+\Delta B\\
&+\Delta b_{lc}\cos^2\theta
\frac{J^2(\sin\theta -1)(4\Omega^2+J^2\sin\theta)}{2(J^2-4\Omega^2)(J^2\sin^2\theta-4\Omega^2)},
        \label{eq:Hsys}
    \end{split}
\end{equation}
with the orbital mixing angle $\theta = \arctan[(\delta+J/2)/2t_{lc}]$.
To predict cavity transmissions in \cref{sec:inputoutput} for  $\omega_c\gtrsim 15~\mu \rm{eV}\approx 3.6\,{\rm GHz}$,
second-order energy corrections have to be considered, see \cref{sec:apsecond}.
In addition to the renormalized qubit parameters, we find an effective 
transversal spin-photon coupling $g_x$ 
and a longitudinal 
effective spin photon coupling $g_z$, both of the order $g_c/\Omega$,
\begin{equation}
    \begin{split}
g_x=&-\frac{\Delta b_{lc}}{2}g_c\cos(\theta)^2 \\
& \times\left(\frac{4\Omega}{J^2-4\Omega^2}
  -\frac{4\Omega}{(2\omega_c+J\sin(\theta))^2
  -4\Omega^2}\right),\\
g_z=&\frac{1}{8}Jg_c\cos(\theta)^2\left(\frac{4\Omega}{4\Omega^2-(2\omega_c+J\sin(\theta))^2}\right.\\
& \left.+\frac{4\Omega}{4\Omega^2-(2\omega_c-J\sin(\theta))^2}
+\frac{8\Omega}{4\Omega^2-J^2\sin(\theta)^2}\right).
        \label{eq:couplingseo}
    \end{split}
\end{equation}
Unlike the flopping-mode qubit that comprises only 
transversal coupling in its proposed operating regime \cite{BenitoFloppingmode2_NoiseTheoExpPhysRevB.100.125430}, 
the singlet-triplet flopping-mode qubit thus inherits the 
observed ability of the ST qubit to couple longitudinally~\cite{UngererbottcherSTlong2022parametric}.
We predict strong longitudinal couplings with the possibility that $g_z > g_x$  
in the operating regime of our model, thus
exceeding the transversal coupling. 
Both couplings are plotted in \cref{fig:cavitycouplings}.
The couplings are highly tunable, with $g_x/g_c,~g_z/g_c \rightarrow 0$ 
for $\delta \gg |t_{lc}|$.
%with
%$(g_{x,z}/g_c)_\text{max}\approx 10 \,( g_{x,z}/%g_c)_\text{min}$ for $\delta \in \{0,200\}$.
%
%///////////////////////////////////////////////////////////////////////
%

%///////////////////////////////////////////////////////////////////////
%

%+++++++++++++++++++++++++++++++++++++++++++++++++++++++++++++++++
\section{Cavity transmission}
\label{sec:inputoutput}
%+++++++++++++++++++++++++++++++++++++++++++++++++++++++++++++++++

\subsection{Input-output theory}
\label{sec:inputoutputsub}

To extract the couplings from the cavity transmission
we use input-output theory \cite{Gardiner1985,burkard2020superconductor_EDyn}. 
The goal is to calculate 
the cavity transmission $T = |\braket{a_{{\rm out},2}}/\braket{a_{{\rm in},1}}|^2$ in the quasistatic limit,
thus the ratio of the output signal
$\braket{a_{{\rm out},2}}$ at port 2 and the input signal
of the probe field at port 1, $\braket{a_{{\rm in},1}}$.
The output signal is related to the cavity field via $a_{{\rm out},2}=\sqrt{\kappa_2}a$.
%So following \cref{fig:thesetupcavityst0}
%we want to calculate 
%$T = \bigl| \frac{\langle a_{out,2}\rangle}{\langle a_{in,1}\rangle}\bigr|^2$.
For this, we first diagonalize the $2\times 2$ effective qubit Hamiltonian
\begin{align}
    H_\text{qb} = \frac{J_\text{eff}}{2} \sigma_z + \frac{b_\text{eff} }{2}\sigma_x,
    \label{eq:Hqubiteff}
\end{align}
and obtain the qubit energies
\begin{align}
    E_{0,1}=\pm \frac{1}{2}\sqrt{J_\text{eff} ^2+ b_\text{eff}^2},
\end{align}
with $E_1 >0>E_0$.
Next, we solve the quantum Langevin equations for the
system in the stationary limit, i.e., $t \rightarrow \infty$,
such that 
$\dot a=\dot\sigma_{ij}=0$, for $i,j\in \{0,1\}$,
with $\sigma_{ij}=\ket{i}\bra{j}$. 
Here, $\ket{i},~i\in \{0,1\}$ is the $i$th eigenstate of the effective qubit Hamiltonian~\eqref{eq:Hqubiteff}. 
Applying a rotating wave approximation to 
remove the time dependence in $H$ arising from the probe field, these equations become
\begin{align}
\begin{split}
      0 =&\dot a =-i[a,H_{\rm{eff}}]\\ &=-i\left(\Delta_c a -  g_{01}\sigma_{01}e^{i\omega_r t}\right)-\frac{\kappa}{2}a+\sqrt{\kappa_{1}}a_{{\rm in},1},
      \end{split}
      \label{eq:atime}
\end{align}
and
\begin{align}
\begin{split}
    0=&\dot{\tilde{\sigma}}_{01} = -i[\tilde\sigma_{01},H_{\rm{eff}}] \\
    &= -i\left(\left(E_1-E_0-\omega_r -i\gamma\right)\tilde\sigma_{01}+g_{01}\left(p_0-p_1\right)a\right),
    \end{split}
    \label{eq:s01}
\end{align}
with $\tilde{\sigma}_{01} =\sigma_{01}e^{i\omega_r t}$, $\Delta_c =\omega_r-\omega_c$, the thermal population 
$p_i=\braket{\sigma_{ii}}=\exp (-E_i/k_B T_{\rm{dot}})$, $i=0,1$, and
\begin{align}
    g_{01} = g_x \frac{J_{\rm eff}}{\sqrt{J_{\rm eff}^2+b_{\rm eff}^2}} - g_z\frac{b_{\rm eff}}{\sqrt{J_{\rm eff}^2+b_{\rm eff}^2}}.
    \label{eq:g01eq}
\end{align}
In the following, we omit the tilde and set $T_{\rm{dot}}=0.75~\rm{K}$.
Considering the expectation values of $a$ and $\sigma_{01}$ 
we can rewrite \cref{eq:s01} as 
\begin{equation}
    \begin{gathered}
   \langle \sigma_{01}\rangle = \chi (\omega_{r})\langle a \rangle,~~\chi(\omega_{r})=\frac{-g_{01}(p_{0}-p_{1})}{E_{1}-E_{0}-\omega_{r}-i\gamma/2}.
    \label{eq:s01exp}
\end{gathered}
\end{equation}
The function $\chi$
is the electric susceptibility of the system~\cite{burkard2020superconductor_EDyn}, which 
encodes the qubit physics in the cavity transmission.
In the following, we will neglect qubit dephasing and set $\gamma=0$.
Substituting this into \eqref{eq:atime} and again replacing $a$
with its quasi-static expectation value, we find
\begin{gather}
   T= \left|\frac{-i\sqrt{\kappa_{1}\kappa_{2}}}{\omega_c-\omega_r+g_{01}\chi-i\kappa/2}\right|^2.
    \label{eq:cavitytrans}
\end{gather}
%
%
%

%
%///////////////////////////////////////////////////////////////////////
\begin{figure}[t]
	\centering
	\includegraphics[width=0.8\linewidth]{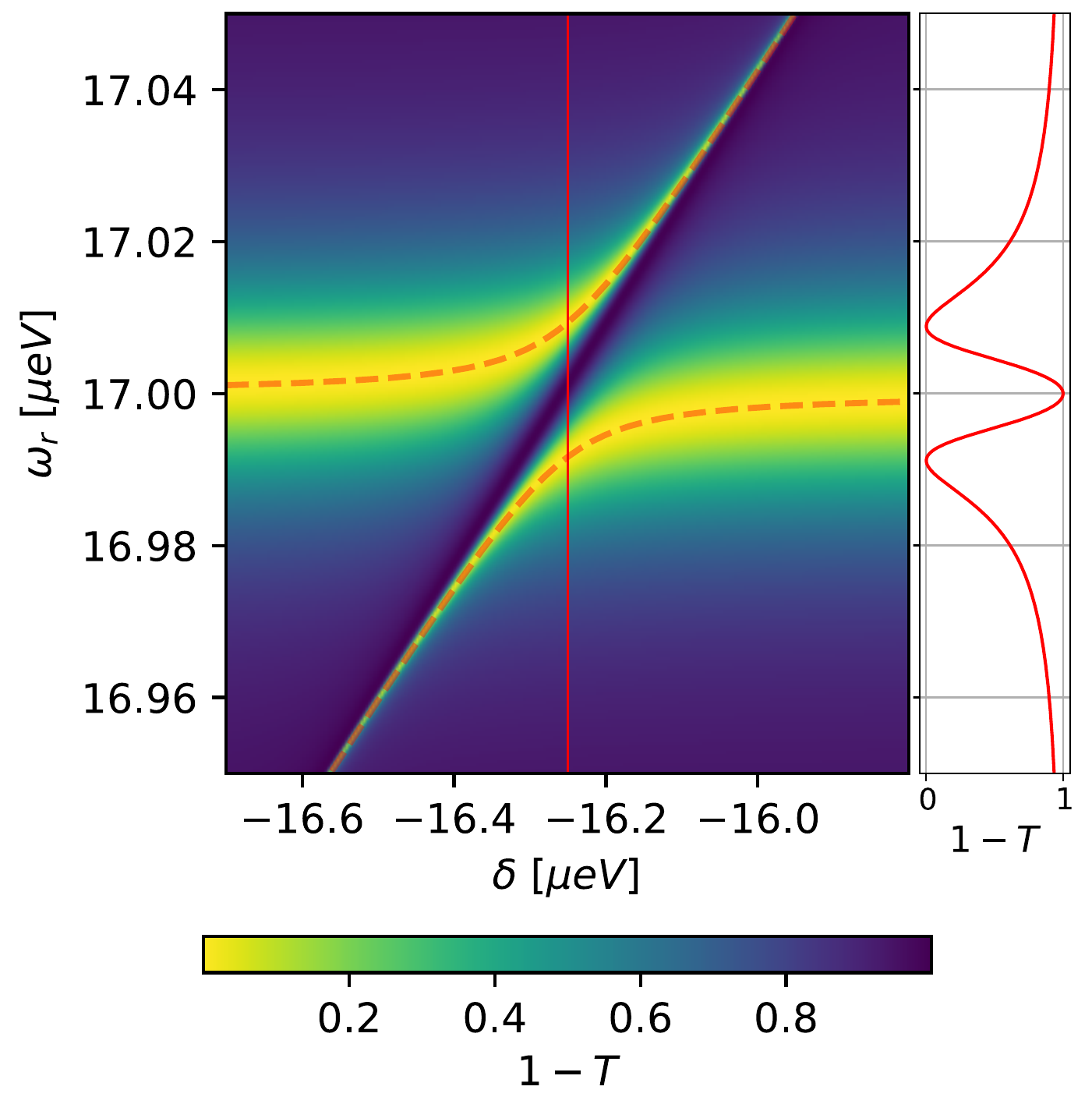}
	\caption{%
	\emph{Calculated cavity absorption $1-T$ as a function of detuning $\delta$ and probe frequency $\omega_r$.}
    Note the avoided crossing around the qubit resonance at $\delta=-16.257~\mu {\rm eV}\approx -J/2 $.
    The absorption spectrum is cut through the avoided crossing
    along the vertical red line to determine $1-T(\omega_r)$, see the right panel. 
    The distance between the two peaks indicates the vacuum Rabi splitting. 
    The width of
    the avoided crossing is $2\lvert g_{01}\rvert\sqrt{p_0-p_1}=0.018~\mu {\rm eV}$, resulting in
    $g_{01}=-0.025~\mu {\rm eV}$, with $p_0=0.5653$ and $p_1=0.4346$.
    Parameters for the plot are $t_{lc}=50~\mu {\rm eV}$, $J=33~\mu {\rm eV}$,
    $B_{l,z}=7~\mu {\rm eV}$, $B_{c,z}=5~\mu {\rm eV}$ and $B_{r,z}=2~\mu {\rm eV}$. The resonator frequency is chosen as $\omega_c = 17~\mu {\rm eV} = 4.1\,{\rm GHz}$.
    }
	\label{fig:cavitytransmission}
\end{figure}
%
%
%///////////////////////////////////////////////////////////////////////
%
%
%
%+++++++++++++++++++++++++++++++++++++++++++++++++++++++++++++++++
\subsection{Results}
\label{sec:results}
%+++++++++++++++++++++++++++++++++++++++++++++++++++++++++++++++++

Both $g_x$ and $g_z$ can be extracted from the cavity transmission $T$ by modulating the frequency of the drive field $\omega_r$. 
At the resonance frequency $\omega_r=\omega_c$ we obtain absorption unless the qubit energy splitting is on resonance with the cavity,
that is $\Delta E = E_1-E_0=\omega_c$. Then we find
an avoided crossing in the cavity spectrum, and the minimal width
of the crossing (see \cref{fig:cavitytransmission}) is $2g_{01}\sqrt{p_0-p_1}$.
The absorption $1-T$ and a cross-section of the 
spectrum around the resonance of the 
qubit at $\delta+J/2 = -16.257~\mu {\rm eV}$ are depicted in \cref{fig:cavitytransmission} for
$t_{lc}=50~\mu {\rm eV}$, $J=33~\mu {\rm eV}$,
$B_{l,z}=7~\mu {\rm eV}$, $B_{c,z}=5~\mu {\rm eV}$ and $B_{r,z}=2~\mu {\rm eV}$. The resonator frequency is chosen as $\omega_c/2\pi = 17\,\mu {\rm eV}=4.1\,{\rm GHz}$. For $T=0.75~\rm{K}$ we have $p_0=0.565$ 
and $p_1=0.435$.
For these parameters, we calculate
$g_{01} = 0.018~\mu {\rm eV}$.  
The coupling $g_{01}$ depends on $g_x$ and on $g_z$, 
see \cref{eq:g01eq},
and to extract these coupling parameters individually from the cavity output,
a single cavity measurement is not sufficient. 
However, one can obtain $g_x$ and $g_z$ in two ways. 
One possibility is using 
that $g_x=0$ for
$\Delta b =0$, so $g_z|_{\Delta b =0}$ can be
resolved. By tuning $\Delta b$,
one can then use the resolved $g_z$ and
calculate $g_x$
from the measurement. Since $\Delta b$ can usually not be tuned in situ, 
this would require measurements across several devices 
and thus demands an otherwise very high similarity of the devices.  
Another way of resolving both couplings is by measuring 
the cavity phase $\phi=\arg(\braket{a_{{\rm out},2}}/\braket{a_{{\rm in},1}})$ in addition to the transmission, 
see \cref{fig:cavityphasest0} in \cref{sec:appphas}, for
the same parameters. 
We calculate $g_x/g_c = -0.010$ 
and $g_z/g_c = 0.168$ for this case. 

The couplings for different $J$ are plotted as a function of $\delta$ 
in \cref{fig:cavitycouplings}.
The two couplings can, 
in a parameter regime as the used one, have opposite signs. At the resonance 
$\delta-J/2=0$, the two charge states are degenerate, and we find,
as for the flopping-mode qubit \cite{PhysRevB.96.235434}, maximal
spin-photon couplings at this operating point. 
%
%
%////////////////////////////////////////////////
\begin{figure}[t]
	\centering
	\includegraphics[width=0.85\linewidth]{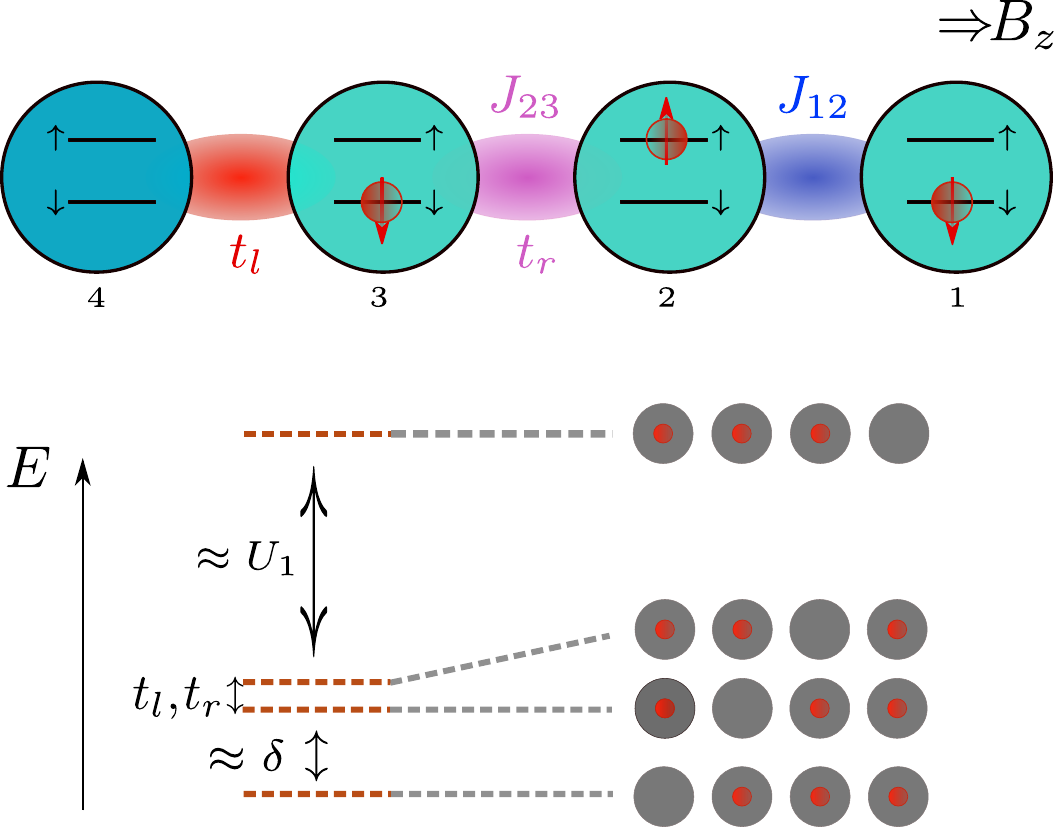}
	\caption{\emph{Exchange-only flopping-mode spin qubit.} 
    Top: Four QDs in a global magnetic field $B_z$.
    Dots 1, 2, and 3 form the EO qubit when dot 4 is far detuned. The 
    exchange couplings $J_{ij}$ between dots $i$ and $j$ and the 
    tunnel couplings $t_l$ between dots 3 and 4 
    and $t_r$ between dots 2 and 3 control the qubit-cavity coupling and the qubit operation.
    Bottom: The four possible charge states without double occupation, organized according to their energy $E$. 
    The state $(1,1,1,0)$ is gapped out by $U_1 \gg \delta$. If $\delta \gg |t_l|, |t_r|$,
    the ground state is an EO qubit coupled to an empty dot and
    if $\delta \ll |t_l|$, the three low-lying charge states hybridize,
    enabling the spin-photon coupling.
    }
	\label{fig:exo_model}
\end{figure}
%////////////////////////////////////////////////
%

%+++++++++++++++++++++++++++++++++++++++++++++++++++++++++++++++++
\section{Exchange-only Flopping-mode Qubit}
\label{sec:floppingexchange}
%+++++++++++++++++++++++++++++++++++++++++++++++++++++++++++++++++

We now apply our 
flopping-mode approach to the exchange-only (EO) qubit. 
The EO qubit can be realized with three QDs and three
electrons, one in each dot~\cite{divincenzo2000universal,
FongWandzureEOquantumcomputingDFSubspace}. 
Analogously to the $ST_0$ qubit, we study the case where 
an additional, empty QD is tunnel coupled to the EO qubit.
The treatment of the system is similar to 
the $ST_0$ case, thus we only highlight the distinct 
aspects in detail. We now 
have a model with four QDs, labeled 1 through 4, occupied with three electrons,
see \cref{fig:exo_model}. A global magnetic field is applied
to separate the two decoherence-free subspaces of the 
EO qubit (see below). The relevant independent parameters are now the exchange couplings
$J_{12}$ [$J_{23}$] between dots (1) and (2) [(2)and (3)]
and the tunnel coupling $t_l$ between dots 3 and 4. The tunnel coupling $t_r$ between dots 2 and 3,
 and the exchange interaction $J_{34}$ between dots 3 and 4, are related to these quantities via 
$J_{23}\propto t_r^2$ and $J_{34}\propto t_l^2$, see
\cref{fig:exo_model}. 
As the $ST_0$ case, our EO model neglects 
high-energy terms due to onsite Coulomb repulsion of 
doubly-occupied dots $U_{2j}$, $j=1,2,3,4$. 
While the $ST_0$ flopping-mode qubit comprises one additional low-energy charge state, we find that the EO flopping-mode qubit comes with two additional charge states.
The four low-energy charge
configurations are $(1,1,1,0), (1, 1,0,1), (1,0,1,1),$ and $(0,1,1,1)$,
see \cref{fig:exo_model}.
The configuration $(1,1,1,0)$ can be neglected, as in the case of the $(1,1,0)$ charge state in the $ST_0$ flopping-mode qubit, because it is gapped out by the large nearest-neighbor Coulomb energy $U_1$. Thus, three states remain, and if dot 4 is detuned by $\delta_{{\rm eo}}$,
then the Hamiltonian of the flopping-mode EO qubit including charge and spin degrees of freedom can then be expressed as
\begin{equation}
\begin{split}
	H_{{\rm fm}, 3}=& \left[J_{12}\bm S_{1}\cdot\bm S_{2}+J_{23}\bm S_{2}\cdot\bm S_{3}	
	-\frac{J_{12}+J_{23}}{4}\right]\\
    &\quad\times \ket{0111}\bra{0111}\\
	&+\left[\delta_{\rm{eo}}+J_{12}\bm S_{1}\cdot\bm S_{2}-\frac{J_{12}}{4}\right]\ket{1011}\bra{1011}\\
	&+\left[\delta_{\rm{eo}}+J_{34}\bm S_{2}\cdot\bm S_{3}-\frac{J_{34}}{4}\right]\ket{1101}\bra{1101}\\
	&+t_{r}\left(\ket{1101}\bra{1011}+h.c.\right) \\
    &+ t_{l}\left(\ket{1011}\bra{0111}+h.c.\right) ,
\end{split}
\label{eq:exchange-only-ham}
\end{equation}
where $\bm S_i = \frac{1}{2}\bm \sigma_i$ is the electron spin operator for electron $i$. 
The computational spin subspace of the EO qubit is spanned by the states
$\ket{S_{123}=\frac{1}{2},S_{12}=0,m},$ and 
$\ket{S_{123}=\frac{1}{2},S_{12}=1,m}$, where $m=-1/2,1/2$ denotes the projection of the total spin of the three particles
$S_{123}$
to the $z$-axis, and $S_{12}= 0,1$ is the total spin
quantum number of the first two electrons.
The qubit states can be written as
\begin{align}
    S_{12}&=0:~\ket{0}=\ket{S}\ket{m},\\
    S_{12}&=1:~\ket{1}=\frac{1}{\sqrt{3}}\left(\sqrt{2}\ket{T_{2m}}\ket{-m}-\ket{T_0}\ket{m}\right),
\end{align}
where $\ket{T_{2m=1}} = \ket{T_+}=\ket{\uparrow \uparrow}$ and $\ket{T_{2m=-1}} = \ket{T_-}=\ket{\downarrow \downarrow}$ are the polarized triplet states of the first two spins. Each of the two subspaces with either $m=+1/2$ or $m=- 1/2$ realizes a valid qubit encoding, however, they are degenerate in energy unless a global Zeeman
field is applied to separate them.
Because \cref{eq:exchange-only-ham} does not mix this subspace 
with the remaining eigenstates, we restrict the Hamiltonian in \cref{eq:exchange-only-ham} to the $m=+1/2$ 
or the $m=-1/2$ subspace and, by adding $(J_{12}+J_{23})/2$ to it, we obtain
\begin{align}
\begin{split}
%\begin{gathered}
%
	&H_\text{eo,comp} =\\
    & \left[\frac{2J_{12}-J_{23}}{4}\sz+\frac{2m\sqrt{3}J_{23}}{4}\sx\right]\ket{0111}\bra{0111}\\
	&+\left[\delta_{\rm{eo}}+\beta+\frac{J_{12}}{2}\sz\right]\ket{1011}\bra{1011}\\
	&+\left[\delta_{\rm{eo}}+\alpha+\frac{J_{34}}{4}(-\sz+2m\sqrt{3}\sx)\right]\ket{1101}\bra{1101}\\
	&+t_{r}(\ket{1101}\bra{1011}+h.c.)
    + t_{l}(\ket{1011}\bra{0111}+h.c.),
\label{eq:exchange-only-ham_compsubspace}
\end{split}
%\end{gathered}
\end{align}
with $\alpha=(J_{12}+J_{23}-J_{34})/2$ and $\beta=J_{23}/2$.
Here, the $\sigma$-Pauli matrices act on the $\ket{0}$, $\ket{1}$ 
states as $\sz\ket{0}=-\ket{0}$ and $\sz\ket{1}=\ket{1}$. 
Note, that the subspaces with $m=\pm 1/2$ differ only by the sign of
the $\sqrt{3}$ terms. In the following analysis, we 
look at $m=+1/2$ (however, see below for a discussion of the subsystem encoding).
The charge degree of freedom of the four dots couples to the electric field of a microwave cavity mode, similarly to the ST qubit (\cref{fig:thesetupcavityst0}) with one additional dot.
The coupling to the cavity field is achieved by coupling dot 4
to the voltage at the center conductor of the cavity. This
varies the energy of the states in which dot 4 is occupied,
which is modeled by,
\begin{equation}
\begin{gathered}
H_{I}=g_c(a+a^\dagger)
(\ket{1011}\bra{1011}+\ket{1101}\bra{1101}).
\label{excoupling}
\end{gathered}
\end{equation}
In addition, a probe field described previously in \cref{eq:Hp} can be injected into the cavity.
Now we can proceed similarly to the singlet-triplet case 
in \cref{sec:model} and derive an effective qubit 
Hamiltonian. However, the situation is more complex
as there is one more charge state to consider.
We can write the energy configuration of the three
charge states as
\begin{align}
\begin{array}{ccc}
&~(1,1,0,1)~~\quad(1,0,1,1)~~\quad (0,1,1,1)&  \\ H_C=
 &
\begin{pmatrix}
 \quad\delta_{\rm{eo}}+\alpha \quad & \quad\quad t_r \quad\quad&\quad\quad 0 \quad\quad \\
\quad\quad t_r\quad \quad& \delta_{\rm{eo}}+\beta & \quad\quad t_l \quad\quad \\
\quad 0 \quad& \quad t_l \quad& \quad\quad0 \quad\quad
\end{pmatrix}. &
\end{array}
\end{align}
The energies of the charge states can be controlled
by $\delta_{\rm{eo}} \ll U_1 \lesssim U_{2j}$ as in the $ST_0$ case. 
If $\delta_{\rm{eo}}+\beta\approx \delta_{\rm{eo}}\gg |t_l|$, the charge
configuration of the ground state is $(0,1,1,1)$. In 
this regime, the system behaves asymptotically like the EO qubit,
i.e., only the $\ket{0111}\bra{0111}$-
terms in the Hamiltonian
\cref{eq:exchange-only-ham_compsubspace} govern the dynamics, thus the system then can be described by
\begin{align}
    &H_\text{eo}= \frac{2J_{12}-J_{23}}{4}\sz+\frac{\sqrt{3}J_{23}}{4}\sx,
\end{align} 
which is the triple-dot EO qubit Hamiltonian \cite{hsieh2012physics}.
For $|\delta_{\rm{eo}}|\ll |t_l|$, the charge states hybridize.
For $|\delta_{\rm{eo}}|\gg |t_l|$, the qubit decouples from the resonator, however, for $\delta_{\rm eo}<0$, as the charge ground state consists of a hybridization of
$(1,0,1,1)$ and $(1,1,0,1)$, 
such that the EO qubit state is not the ground state anymore and becomes unstable.
A suitable idle regime with uncoupled qubit and resonator is therefore $\delta_{\rm eo}>0$ and $\delta_{\rm{eo}}\gg |t_l|$.

\begin{figure}[tb]
	\centering
	\includegraphics[width=0.8\linewidth]{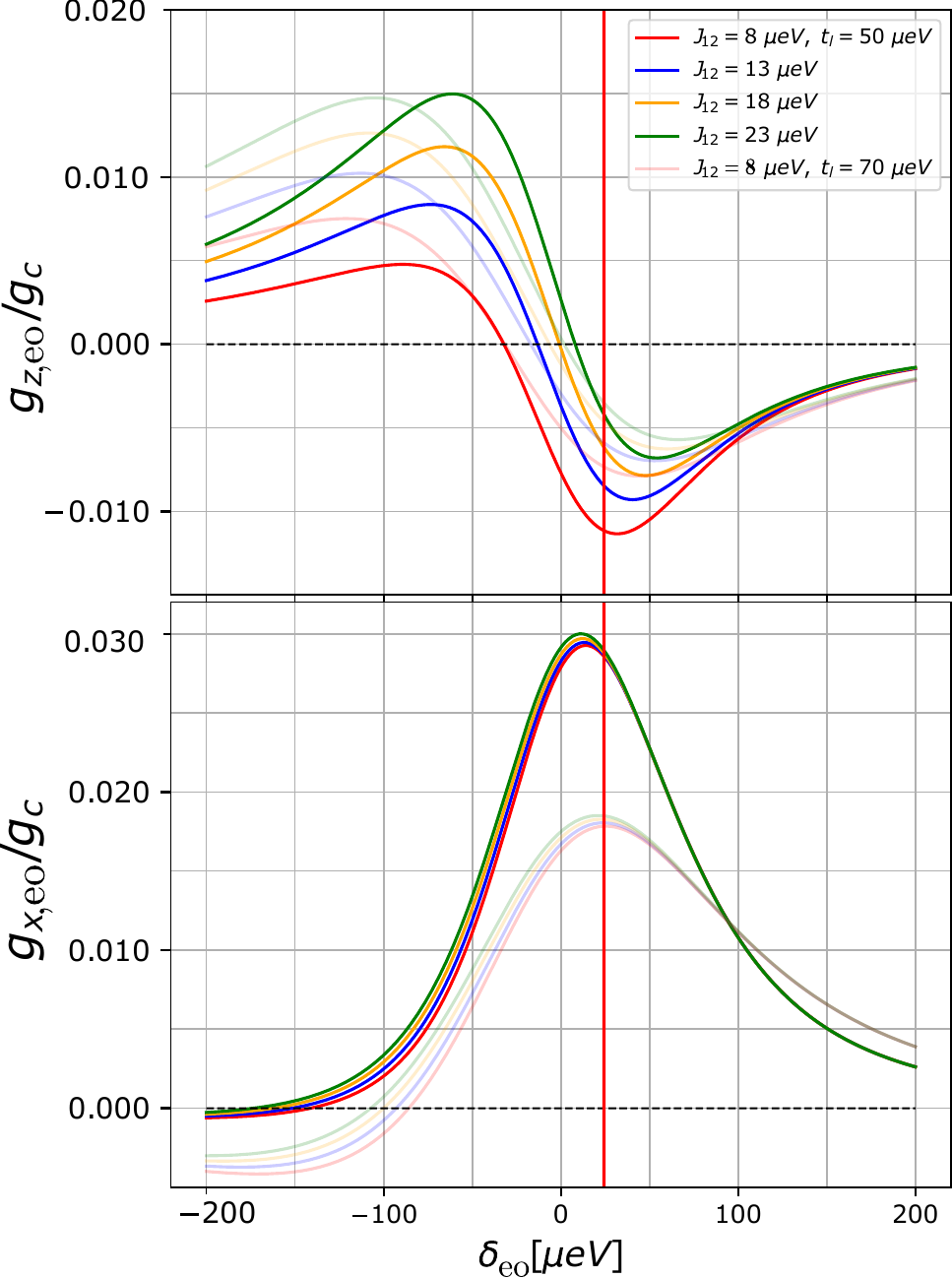}
	\caption{%
	\emph{Cavity couplings for the EO flopping-mode qubit.}
    The effective spin-photon
    couplings $g_{z,\rm{eo}}/g_c$ and $g_{x,\rm{eo}}/g_c$ for $t_{l}=50~\mu \rm{ eV}$, $J_{12}=13~\mu \rm{ eV}$, $t_{r}=25~\mu \rm{ eV}$ (thus $J_{23} = 13.8~\mu \rm{ eV}$) as a function of $\delta_{\rm{eo}}$. The resonator frequency is $\omega_c = 11~\mu \rm{ eV} = 2.66\,{\rm GHz}$. In that regime, the couplings have opposite signs.
    If $-\delta_{\rm{eo}} \approx J_{12}/2$, $g_z$ flips its sign.
    The couplings decays when $\delta_{\rm{eo}} \gg |t_l|$.
    At $\delta_{\rm{eo}}=24.33~\mu \rm{ eV}$ (vertical red line)
    the cavity spectrum is being calculated,
    and there we obtain $g_{z,\rm{eo}}/g_c=-0.0085$ and 
    $g_{x,\rm{eo}}/g_c=-0.029$.
    }
	\label{fig:couplings_exo}
\end{figure}

Now we proceed by diagonalizing the charge sector of 
\cref{eq:exchange-only-ham_compsubspace}, followed by
a Schrieffer-Wolff transformation
in the regime $g_c\ll J_{12},J_{23}\ll |E_i-E_j|$ and $U_{24}\ge U_{2j}$, with $j =1,2,3$, for all eigenvalues $E_i  \neq E_j $ of $H_C$. Here,
the $U_{2j}$ for $j =1,2,3,4$ are the onsite
Coulomb repulsion energies of the four dots.
The resulting effective Hamiltonian is of the form
\begin{equation}
    \begin{split}
H_\text{eff}= & \left[\frac{J_\text{eff,eo}}{2}+g_{z,\rm{eo}}(a+a^\dagger)\right]\tilde\sigma_z\\
&+\left[\frac{j_\text{eff,eo}}{2}+g_{x,\rm{eo}}(a+a^\dagger)\right]\tilde\sigma_x
+\omega_c a^\dagger a.
        \label{eq:Heffeo}
    \end{split}
\end{equation}
%
%Instead, a \textit{adiabatic} ramp up scheme has to be 
%established. The ramp up scheme works as follows:
%start at the decoupled $(0,1,1,1)$ state ($\delta >>t_l,t_r,0$).
%Then ramp up $t_r$ adiabatically. A level crossing 
%occurs if $t_r=\delta$, but
%no scattering between the eigenstates. If $t_r>\delta$,
%$\delta$ can be tuned down and $t_l$ can be tuned up to achieve
%the desired coupling between all charge states. Details are 
%in \todo{Appendix}, however a analysis of this procedure
%with the time dependent Schrödinger equation and the Hamiltonian
%from \cref{eq:exchange-only-ham} still has to be done.
The effective Hamiltonian \cref{eq:Heffeo}, with parameters  given in \cref{sec:appexoparm}, again consists
of the effective qubit Hamiltonian and longitudinal and transversal spin-photon couplings 
$g_{z,\rm{eo}}$ and $g_{x,\rm{eo}}$. 
These couplings are dependent on the tunnel couplings
as well as the detuning $\delta_{\rm{eo}}$. We find that, unlike in the 
$ST_0$ case, the maximal coupling strength is not necessarily
reached where the charge states are degenerate.  Rather, the 
tunnel couplings $t_l$ and $t_r$ can shift the maximum.
The coupling strengths $g_{z,\rm{eo}}$ and $g_{x,\rm{eo}}$ as a function of $\delta_{\rm{eo}}$ are plotted for $m=+1/2$ in \cref{fig:couplings_exo}. The subspace-specific term $2m\sqrt{3}$ appears as a prefactor in $g_x$, while $g_z$ is independent of $m$.
Thus, $g_z$ is identical in both subspaces, and 
$g_x(m=1/2)=-g_x(m=-1/2)$.
We can therefore envision using the longitudinal coupling $g_z$ to couple the \textit{subsystem} qubit to the cavity, and to couple two subsystem qubits via the exchange of cavity photons.

\begin{figure}[t]
	\centering
	\includegraphics[width=0.8\linewidth]{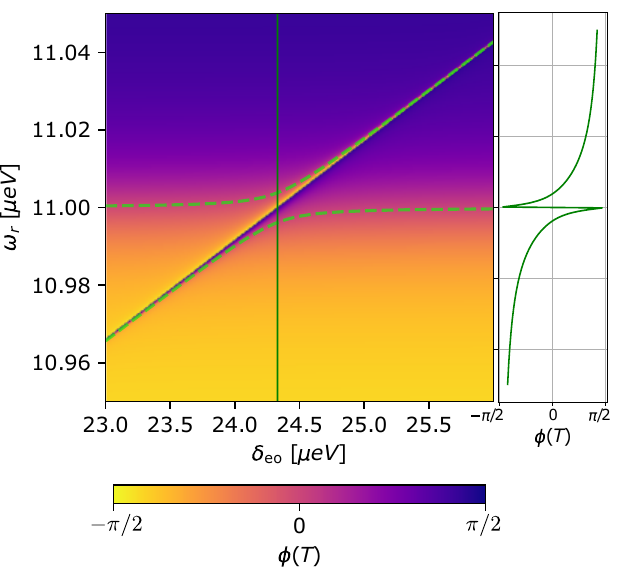}
	\caption{%
	\emph{Cavity phase spectrum for the EO flopping-mode qubit.}
    The cavity phase $\phi$ for $t_{l}=50~\mu \rm{ eV}$, $J_{12}=13~\mu \rm{ eV}$, $t_{r}=25~\mu \rm{ eV}$ (thus $J_{23} \approx 13.8~\mu \rm{ eV}$). The resonator frequency is $\omega_c = 11~\mu \rm{ eV}$, $p_0=0.542$ and $p_1=0.458$.
    Around the qubit resonance at $\delta_{\rm{eo}}=24.33~\mu \rm{ eV}$, 
    we find a $\pi$ phase jump from $-\pi/2$ to $\pi/2$.
    The vertical green line gives $\phi(\omega_r)$ at
    the resonance. The distance of the  
    points with $\phi =0$ next to the 
    phase jump is $2g_{01,\rm{eo}}\sqrt{p_0-p_1}=0.0077~\mu \rm{ eV}$, thus 
    $\lvert g_{01,\rm{eo}}\rvert=0.0132~\mu \rm{ eV}$,
    where $p_0=0.542$ and $p_1=0.458$. 
    }
	\label{fig:cavitytransmission_exo}
\end{figure}

%
%
%///////////////////////////////////////////////////////////////////////
%
%
%+++++++++++++++++++++++++++++++++++++++++++++++++++++++++++++++++
%\subsection{EO: results}
%\label{sec:eoresults}
%+++++++++++++++++++++++++++++++++++++++++++++++++++++++++++++++++
We can use the cavity phase $\phi$
to resolve $g_{x,\rm{eo}}$ and $g_{z,\rm{eo}}$,
as shown in \cref{fig:cavitytransmission_exo} for $t_{l}=50~\mu \rm{ eV}$, $J_{12}=13~\mu \rm{ eV}$, $t_{r}=25~\mu \rm{ eV}$ (thus $J_{23} = 13~\mu \rm{ eV}$). The resonator frequency is set to be 
$\omega_c = 11~\mu \rm{ eV} = 2.66\,{\rm GHz}$. For 
$T=0.75 \rm{K}$ we find $p_0=0.542$ and $p_1=0.458$.
A phase jump of $\pi$ is observed at the resonance.
The two points where $\phi=0$ are separated by 
$\Delta w_r=2g_{01,\rm{eo}}\sqrt{p_0-p_1}$, and 
together with a fit to the cavity transmission 
over $\omega_r$, shown in 
\cref{fig:cavitytransmission_tramin_exo},
$g_{x,\rm{eo}}$ and $g_{z,\rm{eo}}$ 
are extracted. For the given parameters, we find $2\lvert g_{01,\rm{eo}}\rvert\sqrt{p_0-p_1}=0.0078~\mu \rm{ eV}$, thus 
$ g_{01,\rm{eo}}=0.013~\mu \rm{ eV}$. 
The cavity transmission, which is plotted 
in \cref{fig:cavitytransmission_tramin_exo} in 
\cref{sec:appphas}, can also be used to obtain 
$g_{01,\rm{eo}}$.
Calculating $g_{01,\rm{eo}}$ from the effective parameters 
of \cref{eq:Heffeo} gives $g_{01,\rm{eo}}= 0.013~\mu \rm{ eV}$,
and this results in $g_{z,\rm{eo}}/g_c=-0.0085$ 
and $g_{x,\rm{eo}}/g_c=0.029$.
We see here, similar to the $ST_0$ flopping-mode qubit, that $g_{z,\rm{eo}}<g_{x,\rm{eo}}$.
From \cref{fig:combparm} in \cref{sec:appexoparm},
we can see that $g_{01,\rm{eo}}=0$
at $\delta_{\rm eo} \approx 90~\mu \rm{eV}$ while
$g_{x,\rm{eo}}$ and $g_{z,\rm{eo}}$ are each nonzero. At this spot,
the qubit decouples from the cavity in the rotating
wave approximation. However, the system is in a hybridized state
of all three charge states. 
Opposing that, in the $\delta_{\rm eo} \gg |t_l|$ regime
in which the ground state charge configuration 
is $(0,1,1,1)$, the EO qubit also decouples, resulting 
in $g_{01,\rm{eo}}=g_{x,\rm{eo}}=g_{z,\rm{eo}}=0$.
The couplings do not vanish for $\delta_{\rm eo}/|t_l| \rightarrow -\infty$. 
This can be explained by observing that in the regime $\delta_{\rm eo}/|t_l| \rightarrow -\infty$, 
the ground state is a hybridization of the 
$(1,0,1,1)$ and $(1,1,0,1)$ charge states,
which has a small residual electric dipole moment, comparable to that of 
the $ST_0$ flopping-mode qubit.
%
%
%///////////////////////////////////////////////////////////////////////
%
%
%///////////////////////////////////////////////////////////////////////
%Comments.............................
%\begin{equation}
%\begin{gathered}
%
%	H_{ex,m}=
%	\left(\sqrt{3}J_{23}\sigma_x-\frac{1}{2}J_{12}(\sigma_z+
%	5\mathds{1})-\frac{J_{12}+J_{23}}{4}\right)\frac{1-\tau_z}{2}
%	\\+
%	\left(-2\sigma_z-\mathds{1}-\frac{J_{12}}{4}\right)
%	\frac{1+\tau_z}{2}.
%
%\label{eq: ex-hamblock}
%\end{gathered}
%\end{equation}
%
%
%
%
%+++++++++++++++++++++++++++++++++++++++++++++++++++++++++++++++++
\section{Summary}
\label{sec:summary}
%+++++++++++++++++++++++++++++++++++++++++++++++++++++++++++++++++
The flopping-mode qubit adds a charge degree of freedom to the LD qubit, to enable the coupling to a 
resonator as well as low-power fast single-qubit gates.
We combine this idea with the ST and EO qubit concepts by proposing to add a tunnel-coupled 
empty QD to the $ST_0$ and the EO qubits. 
We derive an effective Hamiltonian of the emerging singlet-triplet and exchange-only flopping-mode spin qubits.
The resulting spin-photon coupling of these systems
is investigated and 
highly tunable longitudinal and transversal 
couplings are found. By using input-output theory, 
the cavity transmission and phase
are calculated, and the couplings are extracted
from the transmission and phase spectrum.
The tunable interactions allow for electrically 
controllable spin-photon couplings for 
the $ST_0$ and EO spin qubits in the low-energy charge regime.
These represent promising steps to
realize cavity-mediated baseband-controlled 
two-qubit interactions for $ST_0$ and 
EO spin qubits, which are
locally quite challenging to realize.
In addition, the longitudinal coupling is not dispersive and hence not subject to the Purcell effect.
This may increase qubit lifetimes \cite{UngererbottcherSTlong2022parametric}. In
the future, one can extend this analysis and enhance 
other qubits such as the four-spin singlet-only EO qubit to become 
`flopping mode'.
Here we have focused on calculating and extracting 
the qubit-cavity couplings. From here on, one can apply
known properties of the flopping-mode qubit \cite{GinzelreadoutPhysRevB.108.125437,JonasMielkePRXQuantum.2.020347}
to realize a readout protocol and qubit-qubit coupling for the new qubit flavors which are desired for realizing larger quantum arrays
of semiconductor qubits with long-range cavity-mediated couplings. 
%
%
%+++++++++++++++++++++++++++++++++++++++++++++++++++++++++++++++++
\section{Acknowledgments}
\label{sec:Acknowledgments}
%+++++++++++++++++++++++++++++++++++++++++++++++++++++++++++++++++
This work has been supported by the Army Research Office
Grant No.~W911NF-23-1-0104.
%
%\newpage
%\onecolumngrid
\appendix
%
%+++++++++++++++++++++++++++++++++++++++++++++++++++++++++++++++++
\section{Combined coupling parameter}
\label{sec:combparm}
\begin{figure}[b]
	\centering
	\includegraphics[width=0.9\linewidth]{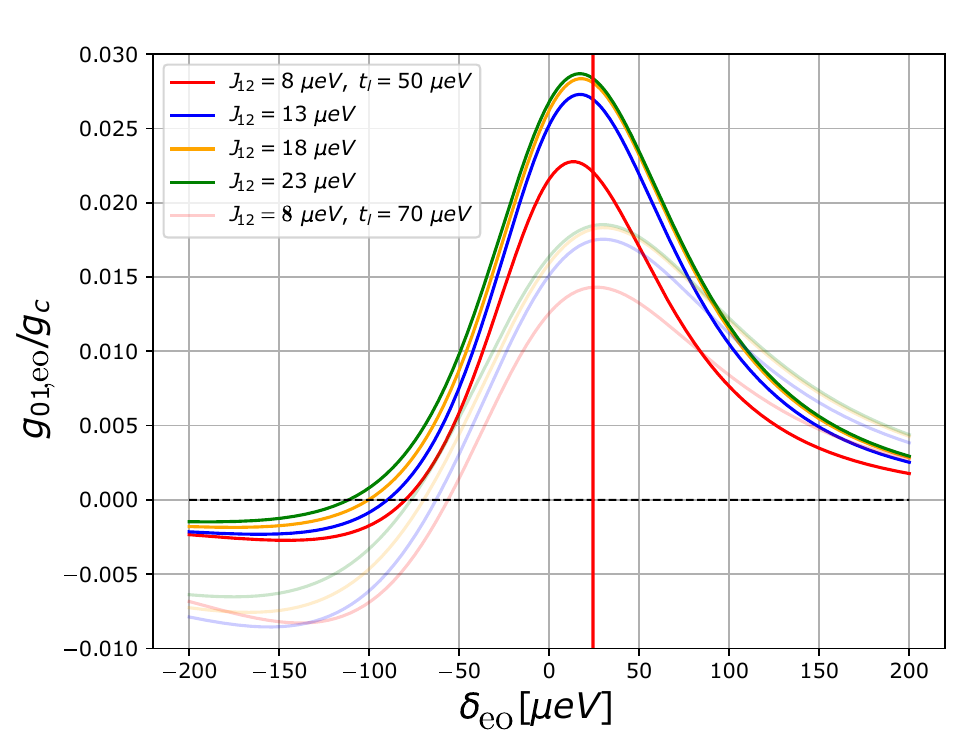}
	\caption{%
	\emph{The coupling parameter $g_{01,\rm{eo}}$ as a function of $\delta_{\rm{eo}}$.}
    The parameters used for this plot are $t_{l}=50~\mu \rm{ eV}$, $J_{12}=13~\mu \rm{ eV}$, $t_{r}=25~\mu \rm{ eV}$ (thus $J_{23} = 13.8~\mu \rm{ eV}$) over $\delta_{\rm{eo}}$. The resonator frequency is $\omega_c = 11~\mu \rm{ eV} = 2.66\,{\rm GHz}$. 
    The coupling decays for $\delta_{\rm{eo}} \gg |t_l|$.
    At $\delta_{\rm{eo}}=24.33~\mu \rm{ eV}$ (vertical red line),
    the cavity spectrum is calculated,
    and we obtain $g_{z,\rm{eo}}/g_c=-0.0085$ and 
    $g_{x,\rm{eo}}/g_c=-0.029$.}
	\label{fig:combparm}
\end{figure}

%+++++++++++++++++++++++++++++++++++++++++++++++++++++++++++++++++
In this section, we define and plot the coupling parameter $g_{01,\rm{eo}}$ (see \cref{fig:combparm}) as it appears in \cref{sec:floppingexchange}, in the cavity transmission and phase for the flopping-mode EO qubit. The coupling $g_{01,\rm{eo}}$ is a combination of the longitudinal and transversal spin-couplings and the effective EO flopping-mode qubit parameters,
\begin{align}
    g_{01,\rm{eo}} = g_{x,\rm{eo}} \frac{J_{\rm eff,eo}}{\sqrt{J_{\rm eff,eo}^2+j_{\rm eff, eo}^2}} - g_{z,\rm{eo}}\frac{j_{\rm eff, eo}}{\sqrt{J_{\rm eff,eo}^2+j_{\rm eff, eo}^2}}.
    \label{eq:g01eqexo}
\end{align}
To decouple the EO qubit from the cavity, $g_{01,\rm{eo}}\rightarrow 0$, one would choose $\delta_{\rm{eo}}\gg |t_l|$.
Note that for typical 
parameters $t_l$, $t_r$, and $J_{12}$, the total coupling $g_{01,\rm{eo}}$ vanishes  around $\delta =-100\,\mu{\rm eV}$. Therefore, within the rotating wave approximation, 
at this point, the qubit can be decoupled from the cavity in addition to the case $\delta/|t_l| \rightarrow \infty$.
However, at the operating point with $\delta<0$, the EO qubit exists in an excited charge state that can decay to a lower-energy state. 
%///////////////////////////////////////////////////////////////////////
%
%
%
%+++++++++++++++++++++++++++++++++++++++++++++++++++++++++++++++++
\section{Transmission phase (modulus) of the $ST_0$ (EO) qubit}
\label{sec:appphas}
%+++++++++++++++++++++++++++++++++++++++++++++++++++++++++++++++++

\begin{figure}
	\centering\includegraphics[width=0.7\linewidth]{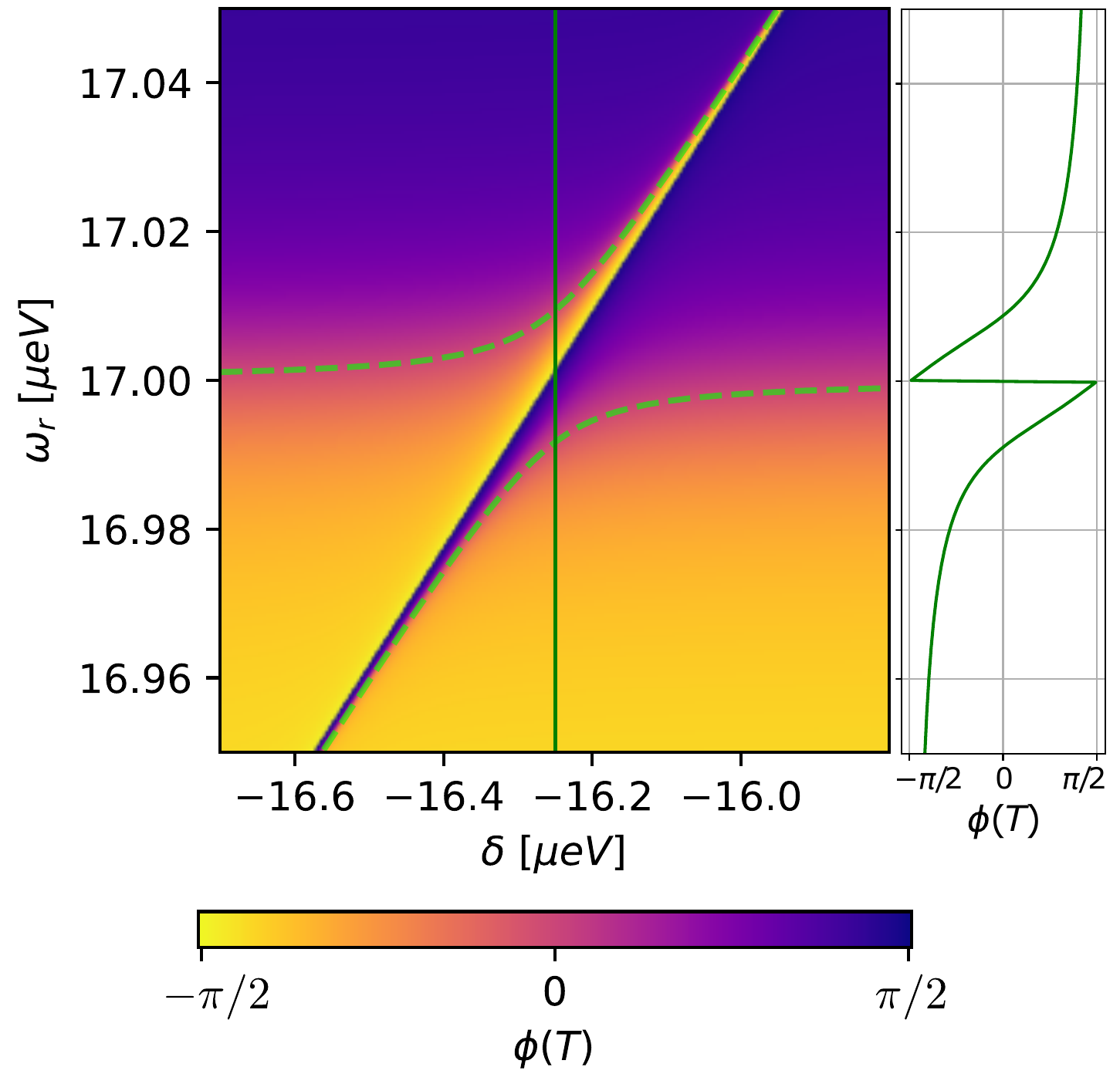}
	\caption{%
	\emph{Cavity phase spectrum of the $ST_0$ 
    flopping-mode qubit.} 
    The cavity phase $\phi$ for $t_{lc}=50~\mu \rm{ eV}$, $J=33~\mu \rm{ eV}$,
    $B_{l,z}=7~\mu \rm{ eV}$, $B_{c,z}=5~\mu \rm{ eV}$ and $B_{r,z}=2~\mu \rm{ eV}$. The resonators frequency is $\omega_c = 17~\mu \rm{ eV}$.
    Around the qubit resonance at $\delta = -16.257~\mu \rm{ eV}$,
    thus $\delta+J/2\approx 0$, we find a 
    $\pi$-phase jump. The spectrum is cut through the avoided crossing
    along the vertical red line to determine $\phi(\omega_r)$. The distance between the two points with
    $\phi=0$ each side of the phase jump 
    is $2\vert g_{01}\vert \sqrt{p_0-p_1}=0.0180~\mu \rm{ eV}$ thus 
    $ g_{01}=-0.0246~\mu \rm{ eV}$. 
    }
	\label{fig:cavityphasest0}
\end{figure}

\begin{figure}
	\centering
\includegraphics[width=0.7\linewidth]{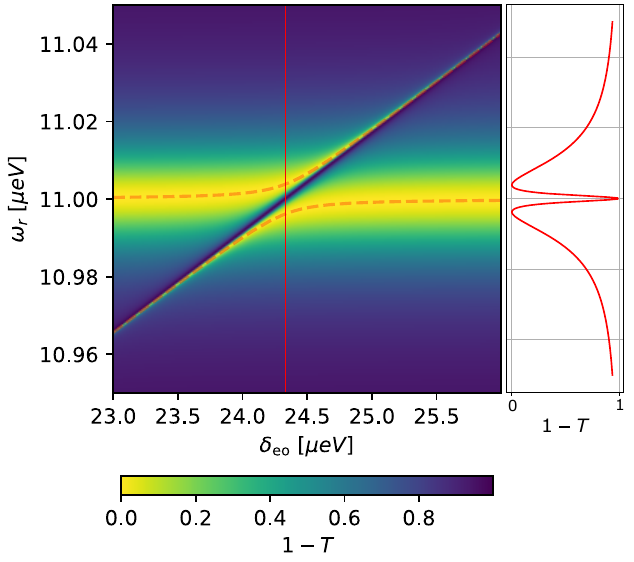}
	\caption{%
	\emph{Cavity absorption spectrum for the EO flopping-mode qubit.} 
      The cavity absorption $1-T$ for $t_{l}=50~\mu \rm{ eV}$, $J_{12}=13~\mu \rm{ eV}$, $t_{r}=25~\mu \rm{ eV}$ (thus $J_{23} \approx 13.8~\mu \rm{ eV}$). The resonator frequency is $\omega_c = 11~\mu \rm{ eV}$.
    Around the qubit resonance at $\delta_{\rm{eo}}=24.33~\mu \rm{ eV}$, 
    we find an avoided crossing.
    The vertical red line gives $1-T(\omega_r)$ at
    the resonance. The width of the avoided crossing is $2g_{01,\rm{eo}}\sqrt{p_0-p_1}=0.0077~\mu \rm{ eV}$ thus 
    $\lvert g_{01,\rm{eo}}\rvert=0.0132~\mu \rm{ eV}$. 
    }
	\label{fig:cavitytransmission_tramin_exo}
\end{figure}

We plot the transmission phase $\phi$ for the $ST_0$ flopping-mode qubit
in \cref{fig:cavityphasest0} for the parameters used in \cref{sec:results}. In \cref{fig:cavitytransmission_tramin_exo}, we plot the absorption $1-T$ 
for the EO flopping-mode qubit
with the parameters used in \cref{sec:floppingexchange}. 
%
%///////////////////////////////////////////////////////////////////////
%
%
%///////////////////////////////////////////////////////////////////////
%
%
%
%
%///////////////////////////////////////////////////////////////////////
%
%
%///////////////////////////////////////////////////////////////////////
%

%\begin{widetext}
%+++++++++++++++++++++++++++++++++++++++++++++++++++++++++++++++++
\section{EO qubit effective parameters}
\label{sec:appexoparm}
%+++++++++++++++++++++++++++++++++++++++++++++++++++++++++++++++++
In this section, we show the parameters of the 
effective Hamiltonian~\cref{eq:Heffeo}. We define
\begin{align}
    \begin{split}
        f&=-(\delta+\alpha)^2+(\delta+\alpha)(\delta+\beta)-(\delta+\beta)^2-3(t_l^2+t_r^2),\\
        g&=2(\delta+\alpha)^3-3(\delta+\alpha)^2(\delta+\beta)-3(\delta+\alpha)(\delta+\beta)^2\\&+2(\delta+\beta)^3-18(\delta+\alpha)t_l^2+9(\delta+\beta)t_l^2+9(\delta+\alpha)t_r^2\\&+9(\delta+\beta)t_r^2,\\
        h&=g+\sqrt{4f^3+g^2}.
    \end{split}
\end{align}
Diagonalizing $H_C$ leads to the charge eigenenergies,
which we can write in a compact form with the cubic roots,
\begin{align}
    \begin{split}
    E_1&=\frac{2\delta+\alpha+\beta}{3}-\frac{2^{1/3}f}{3h^{1/3}}+\frac{h^{1/3}}{3\cdot2^{1/3}},\\
    E_2&=\frac{2\delta+\alpha+\beta}{3}+\frac{1+i\sqrt{3}f}{3\cdot 2^{2/3}h^{1/3}}-\frac{1-i\sqrt{3}h^{1/3}}{6\cdot 2^{1/3}},\\
    E_3&=\frac{2\delta+\alpha+\beta}{3}+\frac{1-i\sqrt{3}f}{3\cdot 2^{2/3}h^{1/3}}-\frac{1+i\sqrt{3}h^{1/3}}{6\cdot 2^{1/3}}.
    \end{split}
\end{align}
%\begin{widetext}
The charge eigenstates are
\begin{align}
    \begin{split}
        \bm v_1 = \left(\frac{-(\delta+\beta)E_1+E_1^2-t_l^2}{t_l t_r},\frac{E_1}{t_l},1\right),\\
        \bm v_2 = \left(\frac{-(\delta+\beta)E_2+E_2^2-t_l^2}{t_l t_r},\frac{E_2}{t_l},1\right),\\
        \bm v_3 = \left(\frac{-(\delta+\beta)E_3+E_3^2-t_l^2}{t_l t_r},\frac{E_3}{t_l},1\right).
    \end{split}
\end{align}
The normalized charge eigenstates are $\bm n_i = \bm v_i/\lvert\bm v_i\rvert$ and for $i,j=1,2,3$ the $j$th component of $\bm n_i$ is denoted as $n_{ij}$. We define
\begin{gather}
    \xi_1=\frac{1}{4} J_{34} \left(1-n_{21}^2\right)-\frac{1}{2} J_{12}
   \left(n_{21}^2+1\right)+\frac{1}{4} J_{23} \left(1-n_{23}^2\right),\\
   \xi_2=\frac{1}{4} J_{34} \left(1-n_{11}^2\right)-\frac{1}{2} J_{12}
   \left(n_{11}^2+1\right)+\frac{1}{4} J_{23} \left(1-n_{13}^2\right),\\
   \xi_3=\frac{J_{34}+2J_{12}}{4}
   \left(n_{31}^2-n_{11}^2\right)+\frac{1}{4} J_{23} \left(n_{33}^2-n_{13}^2\right),\\
   \xi_4=\frac{J_{34}+2J_{12}}{4}
   \left(n_{31}^2-n_{21}^2\right)+\frac{1}{4} J_{23} \left(n_{33}^2-n_{23}^2\right).
\end{gather}
\begin{widetext}
The effective transversal coupling is then,
\begin{align}
\begin{split}
    g_x=&\frac{2m}{8} \sqrt{3} n_{13}n_{33} g_c \bigl(J_{34} n_{11}
   n_{31}+J_{23} n_{13} n_{33}\bigr)\\ 
   &\times \left(\frac{1}{E_1-E_3+\xi_3-\omega
   }-\frac{1}{-E_1+E_3+\xi_3-\omega
   }-\frac{1}{-E_1+E_3+\xi_1}+\frac{1}{E_1-E_3+\xi_1}\right)\\
   +&\frac{2m}{8} \sqrt{3}
   n_{23}n_{33} g_c \left(J_{34} n_{21} n_{31}+J_{23} n_{23}
   n_{33}\right)\\
   & \times\left(\frac{1}{E_2-E_3+\xi_4-\omega
   }-\frac{1}{-E_2+E_3+\xi_4-\omega}-\frac{1}{-E_2+E_3+\xi_2}+\frac{1}{E_2-E_3+\xi_2}\right).
   \end{split}
\end{align}
The effective longitudinal coupling can be written as,
\begin{align}
\begin{split}
    g_z=&\frac{g_c}{16} \left(n_{13} n_{33}\right)  \left(\left(J_{34}+2J_{12}\right) n_{11} n_{31}+J_{23} n_{13} n_{33}\right)
   \left(\frac{2}{
   -E_1+E_3-\xi_3}-\frac{1}{ E_1-E_3+\xi_3-\omega }\right.\\
   &\left.\left.+\frac{1}{-E_1+E_3-\xi_3-\omega }\right)\right.\\
   -&\frac{g_c}{16}
   \left(n_{23} n_{33}\right) \left(n_{21} n_{31} \left(\frac{J_{34}}{4}+\frac{J_{12}}{2}\right)
   + n_{23} n_{33} \frac{J_{23}}{4}\right)
   \left(-\frac{2}{
   -E_2+E_3-\xi_4}+\frac{1}{E_2-E_3+\xi_4-\omega }\right.\\
   &\left.-\frac{1}{-E_2+E_3-\xi_4-\omega }\right)-\left(\xi_i \leftrightarrow - \xi_i,~\omega \leftrightarrow-\omega\right),
   \end{split}
\end{align}
where $(\xi_i \leftrightarrow - \xi_i,~\omega \leftrightarrow-\omega)$ is defined as the same expression again but with $\omega$ and $\xi_i$  replaced by $-\omega$ and $-\xi_i$ for $i =1,2,3,4$. The effective qubit 
parameters are
\begin{align}
    \begin{split}
        j_{\rm eff,eo}=&m\sqrt{3} \left(J_{34} n_{31}^2+J_{23} n_{33}^2\right)+\frac{m\sqrt{3}}{2}\left[ \left(-\frac{1}{4} J_{34} n_{11} n_{31}+\frac{1}{2} J_{12} n_{12}
   n_{32}+\left(\frac{J_{12}}{2}-\frac{J_{23}}{4}\right) n_{13} n_{33}\right)
   \right. \\
   &\left.\times\left(J_{34} n_{11} n_{31}+J_{23} n_{13} n_{33}\right)\left(-\frac{1}{E_1-E_3+\xi_3}+\frac{1}{-E_1+E_3+\xi_1}-\frac{1}{-E_1+E_3+\xi_3}+\frac{1}{E_1-E_3+\xi_1}\right)\right.\\
   &\left.+
   \left(\frac{1}{4} J_{34} n_{21} n_{31}-\frac{1}{2} J_{12} n_{22}
   n_{32}-\left(\frac{J_{12}}{2}+\frac{J_{23}}{4}\right) n_{23} n_{33}\right)\right.\\
   &\left.
   \times\left(J_{34} n_{21} n_{31}+J_{23} n_{23} n_{33}\right) \left(\frac{1}{E_2-E_3+\xi_4}-\frac{1}{-E_2+E_3+\xi_2}+\frac{1}{-E_2+E_3+\xi_4}-\frac{1}{E_2-E_3+\xi_2}\right)
        \right],
    \end{split}
\end{align}
and
\begin{align}
    \begin{split}
        J_{\rm{eff,eo}}=&-\frac{1}{2} J_{34} n_{31}^2+ J_{12}
   n_{32}^2+\left(J_{12}-\frac{J_{23}}{2}\right) n_{33}^2\\
   &+\left[\left(-\frac{1}{2} J_{12} n_{11} n_{31}-\frac{1}{4} J_{34} n_{11}
   n_{31}-\frac{1}{4} J_{23} n_{13} n_{33}\right)^2
   \frac{1}{-E_1+E_3-\xi_3}\right.\\
   &\left.\quad\quad\quad+\frac{3}{16} \left(J_{34} n_{11}
   n_{31}+J_{23} n_{13} n_{33}\right)^2 \frac{1}{-E_1+E_3-\xi_1}\right.\\
   &\left. \quad\quad\quad+ \left(\frac{1}{2} J_{12} n_{21}
   n_{31}+\frac{1}{4} J_{34} n_{21} n_{31}+\frac{1}{4} J_{23} n_{23}
   n_{33}\right)^2 \frac{1}{-E_2+E_3-\xi_4}\right.\\
   &\left.\quad\quad\quad+\frac{3}{16}
   \left(J_{34} n_{21} n_{31}+J_{23} n_{23} n_{33}\right)^2 \frac{1}{-E_2+E_3-\xi_2}
    \quad-\left(\xi_i \leftrightarrow -\xi_i\right)\quad\right].
    \end{split}
\end{align}

%++++++++++++++++++++++++++++++++++++++++++++++++++++
\section{Second order energy contributions to the $ST_0$ flopping-mode qubit}
\label{sec:apsecond}
%+++++++++++++++++++++++++++++++++++++++++++++++++++++++++++++++++
The second-order energy contributions are given as,
\begin{align}
    \begin{split}
        J_{\rm{eff}}^2=\frac{J\cos \theta ^2\left(16 \Delta b_{lc}\left(\Delta b_{lc}-B\right)\Omega^2-\sin \theta\left(4\Delta b_{lc}\Delta BJ^2+J^4+16\Delta b_{lc}^2\Omega^2-4J^2\Omega^2+8\Delta b_{lc}^2J^2\sin \theta\right)\right)}{2\left(J^2-4\Omega^2\right)\left(J^2\sin\theta^2-4\Omega^2\right)},\\
        b_{\rm{eff}}^2=\frac{\cos\theta^2\left(4\left(\Delta b_{lc}-\Delta B\right)J^2\Omega^2+\Delta b_{lc}\sin \theta\left(4\Delta b_{lc}\Delta BJ^2+J^4+16\Delta b_{lc}^2\Omega^2-4J^2\Omega^2-2J^4\sin \theta\right)\right)}{\left(J^2-\Omega^2\right)\left(J^2\sin \theta^2-4\Omega^2\right)}.
    \end{split}
\end{align}

\end{widetext}

%\twocolumngrid
%\bibliographystyle{bibstyle.bst}
\bibliography{biblio.bib}

\begin{thebibliography}{62}%
\makeatletter
\providecommand \@ifxundefined [1]{%
 \@ifx{#1\undefined}
}%
\providecommand \@ifnum [1]{%
 \ifnum #1\expandafter \@firstoftwo
 \else \expandafter \@secondoftwo
 \fi
}%
\providecommand \@ifx [1]{%
 \ifx #1\expandafter \@firstoftwo
 \else \expandafter \@secondoftwo
 \fi
}%
\providecommand \natexlab [1]{#1}%
\providecommand \enquote  [1]{``#1''}%
\providecommand \bibnamefont  [1]{#1}%
\providecommand \bibfnamefont [1]{#1}%
\providecommand \citenamefont [1]{#1}%
\providecommand \href@noop [0]{\@secondoftwo}%
\providecommand \href [0]{\begingroup \@sanitize@url \@href}%
\providecommand \@href[1]{\@@startlink{#1}\@@href}%
\providecommand \@@href[1]{\endgroup#1\@@endlink}%
\providecommand \@sanitize@url [0]{\catcode `\\12\catcode `\$12\catcode
  `\&12\catcode `\#12\catcode `\^12\catcode `\_12\catcode `\%12\relax}%
\providecommand \@@startlink[1]{}%
\providecommand \@@endlink[0]{}%
\providecommand \url  [0]{\begingroup\@sanitize@url \@url }%
\providecommand \@url [1]{\endgroup\@href {#1}{\urlprefix }}%
\providecommand \urlprefix  [0]{URL }%
\providecommand \Eprint [0]{\href }%
\providecommand \doibase [0]{https://doi.org/}%
\providecommand \selectlanguage [0]{\@gobble}%
\providecommand \bibinfo  [0]{\@secondoftwo}%
\providecommand \bibfield  [0]{\@secondoftwo}%
\providecommand \translation [1]{[#1]}%
\providecommand \BibitemOpen [0]{}%
\providecommand \bibitemStop [0]{}%
\providecommand \bibitemNoStop [0]{.\EOS\space}%
\providecommand \EOS [0]{\spacefactor3000\relax}%
\providecommand \BibitemShut  [1]{\csname bibitem#1\endcsname}%
\let\auto@bib@innerbib\@empty
%</preamble>
\bibitem [{\citenamefont {Loss}\ and\ \citenamefont
  {DiVincenzo}(1998)}]{LossDiVincenzoPhysRevA.57.120}%
  \BibitemOpen
  \bibfield  {author} {\bibinfo {author} {\bibfnamefont {D.}~\bibnamefont
  {Loss}}\ and\ \bibinfo {author} {\bibfnamefont {D.~P.}\ \bibnamefont
  {DiVincenzo}},\ }\bibfield  {title} {\bibinfo {title} {Quantum computation
  with quantum dots},\ }\href {https://doi.org/10.1103/PhysRevA.57.120}
  {\bibfield  {journal} {\bibinfo  {journal} {Phys. Rev. A}\ }\textbf {\bibinfo
  {volume} {57}},\ \bibinfo {pages} {120} (\bibinfo {year} {1998})}\BibitemShut
  {NoStop}%
\bibitem [{\citenamefont {Veldhorst}\ \emph {et~al.}(2014)\citenamefont
  {Veldhorst}, \citenamefont {Hwang}, \citenamefont {Yang}, \citenamefont
  {Leenstra}, \citenamefont {de~Ronde}, \citenamefont {Dehollain},
  \citenamefont {Muhonen}, \citenamefont {Hudson}, \citenamefont {Itoh},
  \citenamefont {Morello} \emph
  {et~al.}}]{Veldhorst_coherencetimeslong_veldhorst2014addressable}%
  \BibitemOpen
  \bibfield  {author} {\bibinfo {author} {\bibfnamefont {M.}~\bibnamefont
  {Veldhorst}}, \bibinfo {author} {\bibfnamefont {J.}~\bibnamefont {Hwang}},
  \bibinfo {author} {\bibfnamefont {C.}~\bibnamefont {Yang}}, \bibinfo {author}
  {\bibfnamefont {A.}~\bibnamefont {Leenstra}}, \bibinfo {author}
  {\bibfnamefont {B.}~\bibnamefont {de~Ronde}}, \bibinfo {author}
  {\bibfnamefont {J.}~\bibnamefont {Dehollain}}, \bibinfo {author}
  {\bibfnamefont {J.}~\bibnamefont {Muhonen}}, \bibinfo {author} {\bibfnamefont
  {F.}~\bibnamefont {Hudson}}, \bibinfo {author} {\bibfnamefont {K.~M.}\
  \bibnamefont {Itoh}}, \bibinfo {author} {\bibfnamefont {A.}~\bibnamefont
  {Morello}}, \emph {et~al.},\ }\bibfield  {title} {\bibinfo {title} {An
  addressable quantum dot qubit with fault-tolerant control-fidelity},\ }\href
  {https://rdcu.be/ecbHD} {\bibfield  {journal} {\bibinfo  {journal} {Nature
  {N}anotechnology}\ }\textbf {\bibinfo {volume} {9}},\ \bibinfo {pages} {981}
  (\bibinfo {year} {2014})}\BibitemShut {NoStop}%
\bibitem [{\citenamefont {Tyryshkin}\ \emph {et~al.}(2012)\citenamefont
  {Tyryshkin}, \citenamefont {Tojo}, \citenamefont {Morton}, \citenamefont
  {Riemann}, \citenamefont {Abrosimov}, \citenamefont {Becker}, \citenamefont
  {Pohl}, \citenamefont {Schenkel}, \citenamefont {Thewalt}, \citenamefont
  {Itoh} \emph {et~al.}}]{Coherencetimeslong2_tyryshkin2012electron}%
  \BibitemOpen
  \bibfield  {author} {\bibinfo {author} {\bibfnamefont {A.~M.}\ \bibnamefont
  {Tyryshkin}}, \bibinfo {author} {\bibfnamefont {S.}~\bibnamefont {Tojo}},
  \bibinfo {author} {\bibfnamefont {J.~J.}\ \bibnamefont {Morton}}, \bibinfo
  {author} {\bibfnamefont {H.}~\bibnamefont {Riemann}}, \bibinfo {author}
  {\bibfnamefont {N.~V.}\ \bibnamefont {Abrosimov}}, \bibinfo {author}
  {\bibfnamefont {P.}~\bibnamefont {Becker}}, \bibinfo {author} {\bibfnamefont
  {H.-J.}\ \bibnamefont {Pohl}}, \bibinfo {author} {\bibfnamefont
  {T.}~\bibnamefont {Schenkel}}, \bibinfo {author} {\bibfnamefont {M.~L.}\
  \bibnamefont {Thewalt}}, \bibinfo {author} {\bibfnamefont {K.~M.}\
  \bibnamefont {Itoh}}, \emph {et~al.},\ }\bibfield  {title} {\bibinfo {title}
  {Electron spin coherence exceeding seconds in high-purity silicon},\ }\href
  {https://rdcu.be/ecbHX} {\bibfield  {journal} {\bibinfo  {journal} {Nature
  {M}aterials}\ }\textbf {\bibinfo {volume} {11}},\ \bibinfo {pages} {143}
  (\bibinfo {year} {2012})}\BibitemShut {NoStop}%
\bibitem [{\citenamefont {Wu}\ \emph {et~al.}(2024)\citenamefont {Wu},
  \citenamefont {Camenzind}, \citenamefont {Noiri}, \citenamefont {Takeda},
  \citenamefont {Nakajima}, \citenamefont {Kobayashi}, \citenamefont {Chang},
  \citenamefont {Sammak}, \citenamefont {Scappucci}, \citenamefont {Goan} \emph
  {et~al.}}]{wu2024hamiltonian}%
  \BibitemOpen
  \bibfield  {author} {\bibinfo {author} {\bibfnamefont {Y.-H.}\ \bibnamefont
  {Wu}}, \bibinfo {author} {\bibfnamefont {L.~C.}\ \bibnamefont {Camenzind}},
  \bibinfo {author} {\bibfnamefont {A.}~\bibnamefont {Noiri}}, \bibinfo
  {author} {\bibfnamefont {K.}~\bibnamefont {Takeda}}, \bibinfo {author}
  {\bibfnamefont {T.}~\bibnamefont {Nakajima}}, \bibinfo {author}
  {\bibfnamefont {T.}~\bibnamefont {Kobayashi}}, \bibinfo {author}
  {\bibfnamefont {C.-Y.}\ \bibnamefont {Chang}}, \bibinfo {author}
  {\bibfnamefont {A.}~\bibnamefont {Sammak}}, \bibinfo {author} {\bibfnamefont
  {G.}~\bibnamefont {Scappucci}}, \bibinfo {author} {\bibfnamefont {H.-S.}\
  \bibnamefont {Goan}}, \emph {et~al.},\ }\bibfield  {title} {\bibinfo {title}
  {Hamiltonian phase error in resonantly driven cnot gate above the
  fault-tolerant threshold},\ }\href
  {https://doi.org/10.1038/s41534-023-00802-9} {\bibfield  {journal} {\bibinfo
  {journal} {npj Quantum Information}\ }\textbf {\bibinfo {volume} {10}},\
  \bibinfo {pages} {8} (\bibinfo {year} {2024})}\BibitemShut {NoStop}%
\bibitem [{\citenamefont
  {Kane}(1998)}]{singlet_triplet_Kane1998_kane1998silicon}%
  \BibitemOpen
  \bibfield  {author} {\bibinfo {author} {\bibfnamefont {B.~E.}\ \bibnamefont
  {Kane}},\ }\bibfield  {title} {\bibinfo {title} {A silicon-based nuclear spin
  quantum computer},\ }\href {https://rdcu.be/ecbIj} {\bibfield  {journal}
  {\bibinfo  {journal} {Nature}\ }\textbf {\bibinfo {volume} {393}},\ \bibinfo
  {pages} {133} (\bibinfo {year} {1998})}\BibitemShut {NoStop}%
\bibitem [{\citenamefont {Levy}(2002)}]{Levy2002_PhysRevLett.89.147902}%
  \BibitemOpen
  \bibfield  {author} {\bibinfo {author} {\bibfnamefont {J.}~\bibnamefont
  {Levy}},\ }\bibfield  {title} {\bibinfo {title} {Universal quantum
  computation with spin-$1/2$ pairs and heisenberg exchange},\ }\href
  {https://doi.org/10.1103/PhysRevLett.89.147902} {\bibfield  {journal}
  {\bibinfo  {journal} {Phys. Rev. Lett.}\ }\textbf {\bibinfo {volume} {89}},\
  \bibinfo {pages} {147902} (\bibinfo {year} {2002})}\BibitemShut {NoStop}%
\bibitem [{\citenamefont {Petta}\ \emph {et~al.}(2005)\citenamefont {Petta},
  \citenamefont {Johnson}, \citenamefont {Taylor}, \citenamefont {Laird},
  \citenamefont {Yacoby}, \citenamefont {Lukin}, \citenamefont {Marcus},
  \citenamefont {Hanson},\ and\ \citenamefont
  {Gossard}}]{Petta2005_doi:10.1126/science.1116955}%
  \BibitemOpen
  \bibfield  {author} {\bibinfo {author} {\bibfnamefont {J.~R.}\ \bibnamefont
  {Petta}}, \bibinfo {author} {\bibfnamefont {A.~C.}\ \bibnamefont {Johnson}},
  \bibinfo {author} {\bibfnamefont {J.~M.}\ \bibnamefont {Taylor}}, \bibinfo
  {author} {\bibfnamefont {E.~A.}\ \bibnamefont {Laird}}, \bibinfo {author}
  {\bibfnamefont {A.}~\bibnamefont {Yacoby}}, \bibinfo {author} {\bibfnamefont
  {M.~D.}\ \bibnamefont {Lukin}}, \bibinfo {author} {\bibfnamefont {C.~M.}\
  \bibnamefont {Marcus}}, \bibinfo {author} {\bibfnamefont {M.~P.}\
  \bibnamefont {Hanson}},\ and\ \bibinfo {author} {\bibfnamefont {A.~C.}\
  \bibnamefont {Gossard}},\ }\bibfield  {title} {\bibinfo {title} {Coherent
  manipulation of coupled electron spins in semiconductor quantum dots},\
  }\href {https://doi.org/10.1126/science.1116955} {\bibfield  {journal}
  {\bibinfo  {journal} {Science}\ }\textbf {\bibinfo {volume} {309}},\ \bibinfo
  {pages} {2180} (\bibinfo {year} {2005})}\BibitemShut {NoStop}%
\bibitem [{\citenamefont {DiVincenzo}\ \emph {et~al.}(2000)\citenamefont
  {DiVincenzo}, \citenamefont {Bacon}, \citenamefont {Kempe}, \citenamefont
  {Burkard},\ and\ \citenamefont {Whaley}}]{divincenzo2000universal}%
  \BibitemOpen
  \bibfield  {author} {\bibinfo {author} {\bibfnamefont {D.~P.}\ \bibnamefont
  {DiVincenzo}}, \bibinfo {author} {\bibfnamefont {D.}~\bibnamefont {Bacon}},
  \bibinfo {author} {\bibfnamefont {J.}~\bibnamefont {Kempe}}, \bibinfo
  {author} {\bibfnamefont {G.}~\bibnamefont {Burkard}},\ and\ \bibinfo {author}
  {\bibfnamefont {K.~B.}\ \bibnamefont {Whaley}},\ }\bibfield  {title}
  {\bibinfo {title} {Universal quantum computation with the exchange
  interaction},\ }\href {https://rdcu.be/ecbGl} {\bibfield  {journal} {\bibinfo
   {journal} {Nature}\ }\textbf {\bibinfo {volume} {408}},\ \bibinfo {pages}
  {339} (\bibinfo {year} {2000})}\BibitemShut {NoStop}%
\bibitem [{\citenamefont {Fong}\ and\ \citenamefont
  {Wandzura}(2001)}]{FongWandzureEOquantumcomputingDFSubspace}%
  \BibitemOpen
  \bibfield  {author} {\bibinfo {author} {\bibfnamefont {B.~H.}\ \bibnamefont
  {Fong}}\ and\ \bibinfo {author} {\bibfnamefont {S.~M.}\ \bibnamefont
  {Wandzura}},\ }\bibfield  {title} {\bibinfo {title} {Universal quantum
  computation and leakage reduction in the 3-qubit decoherence free
  subsystem},\ }\href {https://doi.org/https://doi.org/10.26421/QIC11.11-12-9}
  {\bibfield  {journal} {\bibinfo  {journal} {Quantum Information and
  Computation}\ }\textbf {\bibinfo {volume} {11}},\ \bibinfo {pages} {1003}
  (\bibinfo {year} {2001})}\BibitemShut {NoStop}%
\bibitem [{\citenamefont {Bacon}\ \emph {et~al.}(2000)\citenamefont {Bacon},
  \citenamefont {Kempe}, \citenamefont {Lidar},\ and\ \citenamefont
  {Whaley}}]{Baconetal2000_PhysRevLett.85.1758}%
  \BibitemOpen
  \bibfield  {author} {\bibinfo {author} {\bibfnamefont {D.}~\bibnamefont
  {Bacon}}, \bibinfo {author} {\bibfnamefont {J.}~\bibnamefont {Kempe}},
  \bibinfo {author} {\bibfnamefont {D.~A.}\ \bibnamefont {Lidar}},\ and\
  \bibinfo {author} {\bibfnamefont {K.~B.}\ \bibnamefont {Whaley}},\ }\bibfield
   {title} {\bibinfo {title} {Universal fault-tolerant quantum computation on
  decoherence-free subspaces},\ }\href
  {https://doi.org/10.1103/PhysRevLett.85.1758} {\bibfield  {journal} {\bibinfo
   {journal} {Phys. Rev. Lett.}\ }\textbf {\bibinfo {volume} {85}},\ \bibinfo
  {pages} {1758} (\bibinfo {year} {2000})}\BibitemShut {NoStop}%
\bibitem [{\citenamefont {Taylor}\ \emph {et~al.}(2005)\citenamefont {Taylor},
  \citenamefont {Engel}, \citenamefont {D{\"u}r}, \citenamefont {Yacoby},
  \citenamefont {Marcus}, \citenamefont {Zoller},\ and\ \citenamefont
  {Lukin}}]{taylor2005fault}%
  \BibitemOpen
  \bibfield  {author} {\bibinfo {author} {\bibfnamefont {J.}~\bibnamefont
  {Taylor}}, \bibinfo {author} {\bibfnamefont {H.-A.}\ \bibnamefont {Engel}},
  \bibinfo {author} {\bibfnamefont {W.}~\bibnamefont {D{\"u}r}}, \bibinfo
  {author} {\bibfnamefont {A.}~\bibnamefont {Yacoby}}, \bibinfo {author}
  {\bibfnamefont {C.}~\bibnamefont {Marcus}}, \bibinfo {author} {\bibfnamefont
  {P.}~\bibnamefont {Zoller}},\ and\ \bibinfo {author} {\bibfnamefont
  {M.}~\bibnamefont {Lukin}},\ }\bibfield  {title} {\bibinfo {title}
  {Fault-tolerant architecture for quantum computation using electrically
  controlled semiconductor spins},\ }\href {https://rdcu.be/ecbJn} {\bibfield
  {journal} {\bibinfo  {journal} {Nature Physics}\ }\textbf {\bibinfo {volume}
  {1}},\ \bibinfo {pages} {177} (\bibinfo {year} {2005})}\BibitemShut {NoStop}%
\bibitem [{\citenamefont {Taylor}\ \emph {et~al.}(2013)\citenamefont {Taylor},
  \citenamefont {Srinivasa},\ and\ \citenamefont
  {Medford}}]{Taylor_srinivasa_PhysRevLett.111.050502}%
  \BibitemOpen
  \bibfield  {author} {\bibinfo {author} {\bibfnamefont {J.~M.}\ \bibnamefont
  {Taylor}}, \bibinfo {author} {\bibfnamefont {V.}~\bibnamefont {Srinivasa}},\
  and\ \bibinfo {author} {\bibfnamefont {J.}~\bibnamefont {Medford}},\
  }\bibfield  {title} {\bibinfo {title} {Electrically protected resonant
  exchange qubits in triple quantum dots},\ }\href
  {https://doi.org/10.1103/PhysRevLett.111.050502} {\bibfield  {journal}
  {\bibinfo  {journal} {Phys. Rev. Lett.}\ }\textbf {\bibinfo {volume} {111}},\
  \bibinfo {pages} {050502} (\bibinfo {year} {2013})}\BibitemShut {NoStop}%
\bibitem [{\citenamefont {Medford}\ \emph {et~al.}(2013)\citenamefont
  {Medford}, \citenamefont {Beil}, \citenamefont {Taylor}, \citenamefont
  {Rashba}, \citenamefont {Lu}, \citenamefont {Gossard},\ and\ \citenamefont
  {Marcus}}]{Medford_PhysRevLett.111.050501}%
  \BibitemOpen
  \bibfield  {author} {\bibinfo {author} {\bibfnamefont {J.}~\bibnamefont
  {Medford}}, \bibinfo {author} {\bibfnamefont {J.}~\bibnamefont {Beil}},
  \bibinfo {author} {\bibfnamefont {J.~M.}\ \bibnamefont {Taylor}}, \bibinfo
  {author} {\bibfnamefont {E.~I.}\ \bibnamefont {Rashba}}, \bibinfo {author}
  {\bibfnamefont {H.}~\bibnamefont {Lu}}, \bibinfo {author} {\bibfnamefont
  {A.~C.}\ \bibnamefont {Gossard}},\ and\ \bibinfo {author} {\bibfnamefont
  {C.~M.}\ \bibnamefont {Marcus}},\ }\bibfield  {title} {\bibinfo {title}
  {Quantum-dot-based resonant exchange qubit},\ }\href
  {https://doi.org/10.1103/PhysRevLett.111.050501} {\bibfield  {journal}
  {\bibinfo  {journal} {Phys. Rev. Lett.}\ }\textbf {\bibinfo {volume} {111}},\
  \bibinfo {pages} {050501} (\bibinfo {year} {2013})}\BibitemShut {NoStop}%
\bibitem [{\citenamefont {Weinstein}\ \emph {et~al.}(2023)\citenamefont
  {Weinstein}, \citenamefont {Reed}, \citenamefont {Jones}, \citenamefont
  {Andrews}, \citenamefont {Barnes}, \citenamefont {Blumoff}, \citenamefont
  {Euliss}, \citenamefont {Eng}, \citenamefont {Fong}, \citenamefont {Ha} \emph
  {et~al.}}]{weinstein2023universal}%
  \BibitemOpen
  \bibfield  {author} {\bibinfo {author} {\bibfnamefont {A.~J.}\ \bibnamefont
  {Weinstein}}, \bibinfo {author} {\bibfnamefont {M.~D.}\ \bibnamefont {Reed}},
  \bibinfo {author} {\bibfnamefont {A.~M.}\ \bibnamefont {Jones}}, \bibinfo
  {author} {\bibfnamefont {R.~W.}\ \bibnamefont {Andrews}}, \bibinfo {author}
  {\bibfnamefont {D.}~\bibnamefont {Barnes}}, \bibinfo {author} {\bibfnamefont
  {J.~Z.}\ \bibnamefont {Blumoff}}, \bibinfo {author} {\bibfnamefont {L.~E.}\
  \bibnamefont {Euliss}}, \bibinfo {author} {\bibfnamefont {K.}~\bibnamefont
  {Eng}}, \bibinfo {author} {\bibfnamefont {B.~H.}\ \bibnamefont {Fong}},
  \bibinfo {author} {\bibfnamefont {S.~D.}\ \bibnamefont {Ha}}, \emph
  {et~al.},\ }\bibfield  {title} {\bibinfo {title} {Universal logic with
  encoded spin qubits in silicon},\ }\href
  {https://doi.org/10.1038/s41586-023-05777-3} {\bibfield  {journal} {\bibinfo
  {journal} {Nature}\ }\textbf {\bibinfo {volume} {615}},\ \bibinfo {pages}
  {817} (\bibinfo {year} {2023})}\BibitemShut {NoStop}%
\bibitem [{\citenamefont {Ivanova-Rohling}\ \emph {et~al.}(2024)\citenamefont
  {Ivanova-Rohling}, \citenamefont {Rohling},\ and\ \citenamefont
  {Burkard}}]{ivanova2024discovery}%
  \BibitemOpen
  \bibfield  {author} {\bibinfo {author} {\bibfnamefont {V.~N.}\ \bibnamefont
  {Ivanova-Rohling}}, \bibinfo {author} {\bibfnamefont {N.}~\bibnamefont
  {Rohling}},\ and\ \bibinfo {author} {\bibfnamefont {G.}~\bibnamefont
  {Burkard}},\ }\bibfield  {title} {\bibinfo {title} {Discovery of an
  exchange-only gate sequence for cnot with record-low gate time using
  reinforcement learning},\ }\bibfield  {journal} {\bibinfo  {journal} {arXiv
  preprint arXiv:2402.10559}\ }\href
  {https://doi.org/10.48550/arXiv.2402.10559} {10.48550/arXiv.2402.10559}
  (\bibinfo {year} {2024})\BibitemShut {NoStop}%
\bibitem [{\citenamefont {Bose}(2003)}]{BoseSpinBus_PhysRevLett.91.207901}%
  \BibitemOpen
  \bibfield  {author} {\bibinfo {author} {\bibfnamefont {S.}~\bibnamefont
  {Bose}},\ }\bibfield  {title} {\bibinfo {title} {Quantum communication
  through an unmodulated spin chain},\ }\href
  {https://doi.org/10.1103/PhysRevLett.91.207901} {\bibfield  {journal}
  {\bibinfo  {journal} {Phys. Rev. Lett.}\ }\textbf {\bibinfo {volume} {91}},\
  \bibinfo {pages} {207901} (\bibinfo {year} {2003})}\BibitemShut {NoStop}%
\bibitem [{\citenamefont {Bose}(2007)}]{bose2007quantum}%
  \BibitemOpen
  \bibfield  {author} {\bibinfo {author} {\bibfnamefont {S.}~\bibnamefont
  {Bose}},\ }\bibfield  {title} {\bibinfo {title} {Quantum communication
  through spin chain dynamics: an introductory overview},\ }\href
  {https://doi.org/10.1080/00107510701342313} {\bibfield  {journal} {\bibinfo
  {journal} {Contemporary Physics}\ }\textbf {\bibinfo {volume} {48}},\
  \bibinfo {pages} {13} (\bibinfo {year} {2007})}\BibitemShut {NoStop}%
\bibitem [{\citenamefont {Friesen}\ \emph {et~al.}(2007)\citenamefont
  {Friesen}, \citenamefont {Biswas}, \citenamefont {Hu},\ and\ \citenamefont
  {Lidar}}]{friesen2007efficientspinbus}%
  \BibitemOpen
  \bibfield  {author} {\bibinfo {author} {\bibfnamefont {M.}~\bibnamefont
  {Friesen}}, \bibinfo {author} {\bibfnamefont {A.}~\bibnamefont {Biswas}},
  \bibinfo {author} {\bibfnamefont {X.}~\bibnamefont {Hu}},\ and\ \bibinfo
  {author} {\bibfnamefont {D.}~\bibnamefont {Lidar}},\ }\bibfield  {title}
  {\bibinfo {title} {Efficient multiqubit entanglement via a spin bus},\ }\href
  {https://doi.org/10.1103/PhysRevLett.98.230503} {\bibfield  {journal}
  {\bibinfo  {journal} {Phys. Rev. Lett.}\ }\textbf {\bibinfo {volume} {98}},\
  \bibinfo {pages} {230503} (\bibinfo {year} {2007})}\BibitemShut {NoStop}%
\bibitem [{\citenamefont {Sigillito}\ \emph {et~al.}(2019)\citenamefont
  {Sigillito}, \citenamefont {Gullans}, \citenamefont {Edge}, \citenamefont
  {Borselli},\ and\ \citenamefont {Petta}}]{sigillitoPetta2019coherent_bus}%
  \BibitemOpen
  \bibfield  {author} {\bibinfo {author} {\bibfnamefont {A.}~\bibnamefont
  {Sigillito}}, \bibinfo {author} {\bibfnamefont {M.}~\bibnamefont {Gullans}},
  \bibinfo {author} {\bibfnamefont {L.}~\bibnamefont {Edge}}, \bibinfo {author}
  {\bibfnamefont {M.}~\bibnamefont {Borselli}},\ and\ \bibinfo {author}
  {\bibfnamefont {J.}~\bibnamefont {Petta}},\ }\bibfield  {title} {\bibinfo
  {title} {Coherent transfer of quantum information in a silicon double quantum
  dot using resonant swap gates},\ }\href
  {https://doi.org/10.1038/s41534-019-0225-0} {\bibfield  {journal} {\bibinfo
  {journal} {npj Quantum Information}\ }\textbf {\bibinfo {volume} {5}},\
  \bibinfo {pages} {110} (\bibinfo {year} {2019})}\BibitemShut {NoStop}%
\bibitem [{\citenamefont {Craig}\ \emph {et~al.}(2004)\citenamefont {Craig},
  \citenamefont {Taylor}, \citenamefont {Lester}, \citenamefont {Marcus},
  \citenamefont {Hanson},\ and\ \citenamefont {Gossard}}]{craig2004tunable}%
  \BibitemOpen
  \bibfield  {author} {\bibinfo {author} {\bibfnamefont {N.~J.}\ \bibnamefont
  {Craig}}, \bibinfo {author} {\bibfnamefont {J.~M.}\ \bibnamefont {Taylor}},
  \bibinfo {author} {\bibfnamefont {E.~A.}\ \bibnamefont {Lester}}, \bibinfo
  {author} {\bibfnamefont {C.~M.}\ \bibnamefont {Marcus}}, \bibinfo {author}
  {\bibfnamefont {M.~P.}\ \bibnamefont {Hanson}},\ and\ \bibinfo {author}
  {\bibfnamefont {A.~C.}\ \bibnamefont {Gossard}},\ }\bibfield  {title}
  {\bibinfo {title} {Tunable nonlocal spin control in a coupled-quantum dot
  system},\ }\href {https://doi.org/10.1126/science.1095452} {\bibfield
  {journal} {\bibinfo  {journal} {Science}\ }\textbf {\bibinfo {volume}
  {304}},\ \bibinfo {pages} {565} (\bibinfo {year} {2004})}\BibitemShut
  {NoStop}%
\bibitem [{\citenamefont {Malinowski}\ \emph {et~al.}(2019)\citenamefont
  {Malinowski}, \citenamefont {Martins}, \citenamefont {Smith}, \citenamefont
  {Bartlett}, \citenamefont {Doherty}, \citenamefont {Nissen}, \citenamefont
  {Fallahi}, \citenamefont {Gardner}, \citenamefont {Manfra}, \citenamefont
  {Marcus} \emph {et~al.}}]{malinowski2019fast}%
  \BibitemOpen
  \bibfield  {author} {\bibinfo {author} {\bibfnamefont {F.~K.}\ \bibnamefont
  {Malinowski}}, \bibinfo {author} {\bibfnamefont {F.}~\bibnamefont {Martins}},
  \bibinfo {author} {\bibfnamefont {T.~B.}\ \bibnamefont {Smith}}, \bibinfo
  {author} {\bibfnamefont {S.~D.}\ \bibnamefont {Bartlett}}, \bibinfo {author}
  {\bibfnamefont {A.~C.}\ \bibnamefont {Doherty}}, \bibinfo {author}
  {\bibfnamefont {P.~D.}\ \bibnamefont {Nissen}}, \bibinfo {author}
  {\bibfnamefont {S.}~\bibnamefont {Fallahi}}, \bibinfo {author} {\bibfnamefont
  {G.~C.}\ \bibnamefont {Gardner}}, \bibinfo {author} {\bibfnamefont {M.~J.}\
  \bibnamefont {Manfra}}, \bibinfo {author} {\bibfnamefont {C.~M.}\
  \bibnamefont {Marcus}}, \emph {et~al.},\ }\bibfield  {title} {\bibinfo
  {title} {Fast spin exchange across a multielectron mediator},\ }\href
  {https://doi.org/10.1038/s41467-019-09194-x} {\bibfield  {journal} {\bibinfo
  {journal} {Nature Communications}\ }\textbf {\bibinfo {volume} {10}},\
  \bibinfo {pages} {1196} (\bibinfo {year} {2019})}\BibitemShut {NoStop}%
\bibitem [{\citenamefont {Burkard}\ \emph {et~al.}(2023)\citenamefont
  {Burkard}, \citenamefont {Ladd}, \citenamefont {Pan}, \citenamefont
  {Nichol},\ and\ \citenamefont
  {Petta}}]{SpinQubit_Review_RevModPhys.95.025003}%
  \BibitemOpen
  \bibfield  {author} {\bibinfo {author} {\bibfnamefont {G.}~\bibnamefont
  {Burkard}}, \bibinfo {author} {\bibfnamefont {T.~D.}\ \bibnamefont {Ladd}},
  \bibinfo {author} {\bibfnamefont {A.}~\bibnamefont {Pan}}, \bibinfo {author}
  {\bibfnamefont {J.~M.}\ \bibnamefont {Nichol}},\ and\ \bibinfo {author}
  {\bibfnamefont {J.~R.}\ \bibnamefont {Petta}},\ }\bibfield  {title} {\bibinfo
  {title} {Semiconductor spin qubits},\ }\href
  {https://doi.org/10.1103/RevModPhys.95.025003} {\bibfield  {journal}
  {\bibinfo  {journal} {Rev. Mod. Phys.}\ }\textbf {\bibinfo {volume} {95}},\
  \bibinfo {pages} {025003} (\bibinfo {year} {2023})}\BibitemShut {NoStop}%
\bibitem [{\citenamefont {Baart}\ \emph {et~al.}(2016)\citenamefont {Baart},
  \citenamefont {Shafiei}, \citenamefont {Fujita}, \citenamefont {Reichl},
  \citenamefont {Wegscheider},\ and\ \citenamefont
  {Vandersypen}}]{baart2016single}%
  \BibitemOpen
  \bibfield  {author} {\bibinfo {author} {\bibfnamefont {T.~A.}\ \bibnamefont
  {Baart}}, \bibinfo {author} {\bibfnamefont {M.}~\bibnamefont {Shafiei}},
  \bibinfo {author} {\bibfnamefont {T.}~\bibnamefont {Fujita}}, \bibinfo
  {author} {\bibfnamefont {C.}~\bibnamefont {Reichl}}, \bibinfo {author}
  {\bibfnamefont {W.}~\bibnamefont {Wegscheider}},\ and\ \bibinfo {author}
  {\bibfnamefont {L.~M.~K.}\ \bibnamefont {Vandersypen}},\ }\bibfield  {title}
  {\bibinfo {title} {Single-spin {CCD}},\ }\href
  {https://doi.org/10.1038/nnano.2015.291} {\bibfield  {journal} {\bibinfo
  {journal} {Nature {N}anotechnology}\ }\textbf {\bibinfo {volume} {11}},\
  \bibinfo {pages} {330} (\bibinfo {year} {2016})}\BibitemShut {NoStop}%
\bibitem [{\citenamefont {Fujita}\ \emph {et~al.}(2017)\citenamefont {Fujita},
  \citenamefont {Baart}, \citenamefont {Reichl}, \citenamefont {Wegscheider},\
  and\ \citenamefont {Vandersypen}}]{fujita2017coherent}%
  \BibitemOpen
  \bibfield  {author} {\bibinfo {author} {\bibfnamefont {T.}~\bibnamefont
  {Fujita}}, \bibinfo {author} {\bibfnamefont {T.~A.}\ \bibnamefont {Baart}},
  \bibinfo {author} {\bibfnamefont {C.}~\bibnamefont {Reichl}}, \bibinfo
  {author} {\bibfnamefont {W.}~\bibnamefont {Wegscheider}},\ and\ \bibinfo
  {author} {\bibfnamefont {L.~M.~K.}\ \bibnamefont {Vandersypen}},\ }\bibfield
  {title} {\bibinfo {title} {Coherent shuttle of electron-spin states},\ }\href
  {https://doi.org/10.1038/s41534-017-0024-4} {\bibfield  {journal} {\bibinfo
  {journal} {npj Quantum Information}\ }\textbf {\bibinfo {volume} {3}},\
  \bibinfo {pages} {22} (\bibinfo {year} {2017})}\BibitemShut {NoStop}%
\bibitem [{\citenamefont {Burkard}\ \emph {et~al.}(2020)\citenamefont
  {Burkard}, \citenamefont {Gullans}, \citenamefont {Mi},\ and\ \citenamefont
  {Petta}}]{burkard2020superconductor_EDyn}%
  \BibitemOpen
  \bibfield  {author} {\bibinfo {author} {\bibfnamefont {G.}~\bibnamefont
  {Burkard}}, \bibinfo {author} {\bibfnamefont {M.~J.}\ \bibnamefont
  {Gullans}}, \bibinfo {author} {\bibfnamefont {X.}~\bibnamefont {Mi}},\ and\
  \bibinfo {author} {\bibfnamefont {J.~R.}\ \bibnamefont {Petta}},\ }\bibfield
  {title} {\bibinfo {title} {Superconductor--semiconductor hybrid-circuit
  quantum electrodynamics},\ }\href {https://doi.org/10.1038/s42254-019-0135-2}
  {\bibfield  {journal} {\bibinfo  {journal} {Nature Reviews Physics}\ }\textbf
  {\bibinfo {volume} {2}},\ \bibinfo {pages} {129} (\bibinfo {year}
  {2020})}\BibitemShut {NoStop}%
\bibitem [{\citenamefont {Childress}\ \emph {et~al.}(2004)\citenamefont
  {Childress}, \citenamefont {S{\o}rensen},\ and\ \citenamefont
  {Lukin}}]{Longrange1childress2004mesoscopic}%
  \BibitemOpen
  \bibfield  {author} {\bibinfo {author} {\bibfnamefont {L.}~\bibnamefont
  {Childress}}, \bibinfo {author} {\bibfnamefont {A.}~\bibnamefont
  {S{\o}rensen}},\ and\ \bibinfo {author} {\bibfnamefont {M.~D.}\ \bibnamefont
  {Lukin}},\ }\bibfield  {title} {\bibinfo {title} {Mesoscopic cavity quantum
  electrodynamics with quantum dots},\ }\href
  {https://doi.org/10.1103/PhysRevA.69.042302} {\bibfield  {journal} {\bibinfo
  {journal} {Physical Review A}\ }\textbf {\bibinfo {volume} {69}},\ \bibinfo
  {pages} {042302} (\bibinfo {year} {2004})}\BibitemShut {NoStop}%
\bibitem [{\citenamefont {Burkard}\ and\ \citenamefont
  {Imamoglu}(2006)}]{Longrange2burkard2006ultra}%
  \BibitemOpen
  \bibfield  {author} {\bibinfo {author} {\bibfnamefont {G.}~\bibnamefont
  {Burkard}}\ and\ \bibinfo {author} {\bibfnamefont {A.}~\bibnamefont
  {Imamoglu}},\ }\bibfield  {title} {\bibinfo {title} {Ultra-long-distance
  interaction between spin qubits},\ }\href
  {https://doi.org/10.1103/PhysRevB.74.041307} {\bibfield  {journal} {\bibinfo
  {journal} {Physical Review B}\ }\textbf {\bibinfo {volume} {74}},\ \bibinfo
  {pages} {041307} (\bibinfo {year} {2006})}\BibitemShut {NoStop}%
\bibitem [{\citenamefont {B{\o}ttcher}\ \emph {et~al.}(2022)\citenamefont
  {B{\o}ttcher}, \citenamefont {Harvey}, \citenamefont {Fallahi}, \citenamefont
  {Gardner}, \citenamefont {Manfra}, \citenamefont {Vool}, \citenamefont
  {Bartlett},\ and\ \citenamefont
  {Yacoby}}]{UngererbottcherSTlong2022parametric}%
  \BibitemOpen
  \bibfield  {author} {\bibinfo {author} {\bibfnamefont {C.}~\bibnamefont
  {B{\o}ttcher}}, \bibinfo {author} {\bibfnamefont {S.}~\bibnamefont {Harvey}},
  \bibinfo {author} {\bibfnamefont {S.}~\bibnamefont {Fallahi}}, \bibinfo
  {author} {\bibfnamefont {G.}~\bibnamefont {Gardner}}, \bibinfo {author}
  {\bibfnamefont {M.}~\bibnamefont {Manfra}}, \bibinfo {author} {\bibfnamefont
  {U.}~\bibnamefont {Vool}}, \bibinfo {author} {\bibfnamefont {S.}~\bibnamefont
  {Bartlett}},\ and\ \bibinfo {author} {\bibfnamefont {A.}~\bibnamefont
  {Yacoby}},\ }\bibfield  {title} {\bibinfo {title} {Parametric longitudinal
  coupling between a high-impedance superconducting resonator and a
  semiconductor quantum dot singlet-triplet spin qubit},\ }\href
  {https://doi.org/10.1038/s41467-022-32236-w} {\bibfield  {journal} {\bibinfo
  {journal} {Nature Communications}\ }\textbf {\bibinfo {volume} {13}},\
  \bibinfo {pages} {4773} (\bibinfo {year} {2022})}\BibitemShut {NoStop}%
\bibitem [{\citenamefont {Russ}\ \emph {et~al.}(2016)\citenamefont {Russ},
  \citenamefont {Ginzel},\ and\ \citenamefont {Burkard}}]{russ2016coupling}%
  \BibitemOpen
  \bibfield  {author} {\bibinfo {author} {\bibfnamefont {M.}~\bibnamefont
  {Russ}}, \bibinfo {author} {\bibfnamefont {F.}~\bibnamefont {Ginzel}},\ and\
  \bibinfo {author} {\bibfnamefont {G.}~\bibnamefont {Burkard}},\ }\bibfield
  {title} {\bibinfo {title} {Coupling of three-spin qubits to their electric
  environment},\ }\href {https://doi.org/10.1103/PhysRevB.94.165411} {\bibfield
   {journal} {\bibinfo  {journal} {Phys. Rev. B}\ }\textbf {\bibinfo {volume}
  {94}},\ \bibinfo {pages} {165411} (\bibinfo {year} {2016})}\BibitemShut
  {NoStop}%
\bibitem [{\citenamefont {Landig}\ \emph {et~al.}(2018)\citenamefont {Landig},
  \citenamefont {Koski}, \citenamefont {Scarlino}, \citenamefont {Mendes},
  \citenamefont {Blais}, \citenamefont {Reichl}, \citenamefont {Wegscheider},
  \citenamefont {Wallraff}, \citenamefont {Ensslin},\ and\ \citenamefont
  {Ihn}}]{landig2018coherent}%
  \BibitemOpen
  \bibfield  {author} {\bibinfo {author} {\bibfnamefont {A.~J.}\ \bibnamefont
  {Landig}}, \bibinfo {author} {\bibfnamefont {J.~V.}\ \bibnamefont {Koski}},
  \bibinfo {author} {\bibfnamefont {P.}~\bibnamefont {Scarlino}}, \bibinfo
  {author} {\bibfnamefont {U.}~\bibnamefont {Mendes}}, \bibinfo {author}
  {\bibfnamefont {A.}~\bibnamefont {Blais}}, \bibinfo {author} {\bibfnamefont
  {C.}~\bibnamefont {Reichl}}, \bibinfo {author} {\bibfnamefont
  {W.}~\bibnamefont {Wegscheider}}, \bibinfo {author} {\bibfnamefont
  {A.}~\bibnamefont {Wallraff}}, \bibinfo {author} {\bibfnamefont
  {K.}~\bibnamefont {Ensslin}},\ and\ \bibinfo {author} {\bibfnamefont
  {T.}~\bibnamefont {Ihn}},\ }\bibfield  {title} {\bibinfo {title} {Coherent
  spin--photon coupling using a resonant exchange qubit},\ }\href
  {https://doi.org/10.1038/s41586-018-0365-y} {\bibfield  {journal} {\bibinfo
  {journal} {Nature}\ }\textbf {\bibinfo {volume} {560}},\ \bibinfo {pages}
  {179} (\bibinfo {year} {2018})}\BibitemShut {NoStop}%
\bibitem [{\citenamefont {Benito}\ \emph
  {et~al.}(2019{\natexlab{a}})\citenamefont {Benito}, \citenamefont {Croot},
  \citenamefont {Adelsberger}, \citenamefont {Putz}, \citenamefont {Mi},
  \citenamefont {Petta},\ and\ \citenamefont
  {Burkard}}]{BenitoFloppingmode2_NoiseTheoExpPhysRevB.100.125430}%
  \BibitemOpen
  \bibfield  {author} {\bibinfo {author} {\bibfnamefont {M.}~\bibnamefont
  {Benito}}, \bibinfo {author} {\bibfnamefont {X.}~\bibnamefont {Croot}},
  \bibinfo {author} {\bibfnamefont {C.}~\bibnamefont {Adelsberger}}, \bibinfo
  {author} {\bibfnamefont {S.}~\bibnamefont {Putz}}, \bibinfo {author}
  {\bibfnamefont {X.}~\bibnamefont {Mi}}, \bibinfo {author} {\bibfnamefont
  {J.~R.}\ \bibnamefont {Petta}},\ and\ \bibinfo {author} {\bibfnamefont
  {G.}~\bibnamefont {Burkard}},\ }\bibfield  {title} {\bibinfo {title}
  {Electric-field control and noise protection of the flopping-mode spin
  qubit},\ }\href {https://doi.org/10.1103/PhysRevB.100.125430} {\bibfield
  {journal} {\bibinfo  {journal} {Phys. Rev. B}\ }\textbf {\bibinfo {volume}
  {100}},\ \bibinfo {pages} {125430} (\bibinfo {year}
  {2019}{\natexlab{a}})}\BibitemShut {NoStop}%
\bibitem [{\citenamefont {Croot}\ \emph {et~al.}(2020)\citenamefont {Croot},
  \citenamefont {Mi}, \citenamefont {Putz}, \citenamefont {Benito},
  \citenamefont {Borjans}, \citenamefont {Burkard},\ and\ \citenamefont
  {Petta}}]{CrootPhysRevResearch.2.012006}%
  \BibitemOpen
  \bibfield  {author} {\bibinfo {author} {\bibfnamefont {X.}~\bibnamefont
  {Croot}}, \bibinfo {author} {\bibfnamefont {X.}~\bibnamefont {Mi}}, \bibinfo
  {author} {\bibfnamefont {S.}~\bibnamefont {Putz}}, \bibinfo {author}
  {\bibfnamefont {M.}~\bibnamefont {Benito}}, \bibinfo {author} {\bibfnamefont
  {F.}~\bibnamefont {Borjans}}, \bibinfo {author} {\bibfnamefont
  {G.}~\bibnamefont {Burkard}},\ and\ \bibinfo {author} {\bibfnamefont {J.~R.}\
  \bibnamefont {Petta}},\ }\bibfield  {title} {\bibinfo {title} {Flopping-mode
  electric dipole spin resonance},\ }\href
  {https://doi.org/10.1103/PhysRevResearch.2.012006} {\bibfield  {journal}
  {\bibinfo  {journal} {Phys. Rev. Res.}\ }\textbf {\bibinfo {volume} {2}},\
  \bibinfo {pages} {012006} (\bibinfo {year} {2020})}\BibitemShut {NoStop}%
\bibitem [{\citenamefont {Benito}\ \emph
  {et~al.}(2019{\natexlab{b}})\citenamefont {Benito}, \citenamefont {Petta},\
  and\ \citenamefont {Burkard}}]{longbenito2019optimized}%
  \BibitemOpen
  \bibfield  {author} {\bibinfo {author} {\bibfnamefont {M.}~\bibnamefont
  {Benito}}, \bibinfo {author} {\bibfnamefont {J.~R.}\ \bibnamefont {Petta}},\
  and\ \bibinfo {author} {\bibfnamefont {G.}~\bibnamefont {Burkard}},\
  }\bibfield  {title} {\bibinfo {title} {Optimized cavity-mediated dispersive
  two-qubit gates between spin qubits},\ }\href
  {https://doi.org/10.1103/PhysRevB.100.081412} {\bibfield  {journal} {\bibinfo
   {journal} {Physical Review B}\ }\textbf {\bibinfo {volume} {100}},\ \bibinfo
  {pages} {081412} (\bibinfo {year} {2019}{\natexlab{b}})}\BibitemShut
  {NoStop}%
\bibitem [{\citenamefont {Warren}\ \emph {et~al.}(2019)\citenamefont {Warren},
  \citenamefont {Barnes},\ and\ \citenamefont {Economou}}]{warren2019long}%
  \BibitemOpen
  \bibfield  {author} {\bibinfo {author} {\bibfnamefont {A.}~\bibnamefont
  {Warren}}, \bibinfo {author} {\bibfnamefont {E.}~\bibnamefont {Barnes}},\
  and\ \bibinfo {author} {\bibfnamefont {S.~E.}\ \bibnamefont {Economou}},\
  }\bibfield  {title} {\bibinfo {title} {Long-distance entangling gates between
  quantum dot spins mediated by a superconducting resonator},\ }\href
  {https://doi.org/10.1103/PhysRevResearch.6.043029} {\bibfield  {journal}
  {\bibinfo  {journal} {Physical Review B}\ }\textbf {\bibinfo {volume}
  {100}},\ \bibinfo {pages} {161303} (\bibinfo {year} {2019})}\BibitemShut
  {NoStop}%
\bibitem [{\citenamefont {Mi}\ \emph {et~al.}(2017)\citenamefont {Mi},
  \citenamefont {Cady}, \citenamefont {Zajac}, \citenamefont {Deelman},\ and\
  \citenamefont {Petta}}]{CavityQEDmi2017strong}%
  \BibitemOpen
  \bibfield  {author} {\bibinfo {author} {\bibfnamefont {X.}~\bibnamefont
  {Mi}}, \bibinfo {author} {\bibfnamefont {J.}~\bibnamefont {Cady}}, \bibinfo
  {author} {\bibfnamefont {D.}~\bibnamefont {Zajac}}, \bibinfo {author}
  {\bibfnamefont {P.}~\bibnamefont {Deelman}},\ and\ \bibinfo {author}
  {\bibfnamefont {J.~R.}\ \bibnamefont {Petta}},\ }\bibfield  {title} {\bibinfo
  {title} {Strong coupling of a single electron in silicon to a microwave
  photon},\ }\href {https://doi.org/10.1126/science.aal24} {\bibfield
  {journal} {\bibinfo  {journal} {Science}\ }\textbf {\bibinfo {volume}
  {355}},\ \bibinfo {pages} {156} (\bibinfo {year} {2017})}\BibitemShut
  {NoStop}%
\bibitem [{\citenamefont {Hu}\ \emph {et~al.}(2012)\citenamefont {Hu},
  \citenamefont {Liu},\ and\ \citenamefont {Nori}}]{hu2012strong}%
  \BibitemOpen
  \bibfield  {author} {\bibinfo {author} {\bibfnamefont {X.}~\bibnamefont
  {Hu}}, \bibinfo {author} {\bibfnamefont {Y.-x.}\ \bibnamefont {Liu}},\ and\
  \bibinfo {author} {\bibfnamefont {F.}~\bibnamefont {Nori}},\ }\bibfield
  {title} {\bibinfo {title} {Strong coupling of a spin qubit to a
  superconducting stripline cavity},\ }\href
  {https://doi.org/10.1103/PhysRevB.86.035314} {\bibfield  {journal} {\bibinfo
  {journal} {Physical Review B}\ }\textbf {\bibinfo {volume} {86}},\ \bibinfo
  {pages} {035314} (\bibinfo {year} {2012})}\BibitemShut {NoStop}%
\bibitem [{\citenamefont {Mi}\ \emph {et~al.}(2018)\citenamefont {Mi},
  \citenamefont {Benito}, \citenamefont {Putz}, \citenamefont {Zajac},
  \citenamefont {Taylor}, \citenamefont {Burkard},\ and\ \citenamefont
  {Petta}}]{mi2018coherent}%
  \BibitemOpen
  \bibfield  {author} {\bibinfo {author} {\bibfnamefont {X.}~\bibnamefont
  {Mi}}, \bibinfo {author} {\bibfnamefont {M.}~\bibnamefont {Benito}}, \bibinfo
  {author} {\bibfnamefont {S.}~\bibnamefont {Putz}}, \bibinfo {author}
  {\bibfnamefont {D.~M.}\ \bibnamefont {Zajac}}, \bibinfo {author}
  {\bibfnamefont {J.~M.}\ \bibnamefont {Taylor}}, \bibinfo {author}
  {\bibfnamefont {G.}~\bibnamefont {Burkard}},\ and\ \bibinfo {author}
  {\bibfnamefont {J.~R.}\ \bibnamefont {Petta}},\ }\bibfield  {title} {\bibinfo
  {title} {A coherent spin--photon interface in silicon},\ }\href
  {https://doi.org/10.1038/nature25769} {\bibfield  {journal} {\bibinfo
  {journal} {Nature}\ }\textbf {\bibinfo {volume} {555}},\ \bibinfo {pages}
  {599} (\bibinfo {year} {2018})}\BibitemShut {NoStop}%
\bibitem [{\citenamefont {Benito}\ \emph {et~al.}(2017)\citenamefont {Benito},
  \citenamefont {Mi}, \citenamefont {Taylor}, \citenamefont {Petta},\ and\
  \citenamefont {Burkard}}]{PhysRevB.96.235434}%
  \BibitemOpen
  \bibfield  {author} {\bibinfo {author} {\bibfnamefont {M.}~\bibnamefont
  {Benito}}, \bibinfo {author} {\bibfnamefont {X.}~\bibnamefont {Mi}}, \bibinfo
  {author} {\bibfnamefont {J.~M.}\ \bibnamefont {Taylor}}, \bibinfo {author}
  {\bibfnamefont {J.~R.}\ \bibnamefont {Petta}},\ and\ \bibinfo {author}
  {\bibfnamefont {G.}~\bibnamefont {Burkard}},\ }\bibfield  {title} {\bibinfo
  {title} {Input-output theory for spin-photon coupling in si double quantum
  dots},\ }\href {https://doi.org/10.1103/PhysRevB.96.235434} {\bibfield
  {journal} {\bibinfo  {journal} {Phys. Rev. B}\ }\textbf {\bibinfo {volume}
  {96}},\ \bibinfo {pages} {235434} (\bibinfo {year} {2017})}\BibitemShut
  {NoStop}%
\bibitem [{\citenamefont {Borjans}\ \emph {et~al.}(2020)\citenamefont
  {Borjans}, \citenamefont {Croot}, \citenamefont {Mi}, \citenamefont
  {Gullans},\ and\ \citenamefont {Petta}}]{longrange2borjans2020resonant}%
  \BibitemOpen
  \bibfield  {author} {\bibinfo {author} {\bibfnamefont {F.}~\bibnamefont
  {Borjans}}, \bibinfo {author} {\bibfnamefont {X.}~\bibnamefont {Croot}},
  \bibinfo {author} {\bibfnamefont {X.}~\bibnamefont {Mi}}, \bibinfo {author}
  {\bibfnamefont {M.}~\bibnamefont {Gullans}},\ and\ \bibinfo {author}
  {\bibfnamefont {J.}~\bibnamefont {Petta}},\ }\bibfield  {title} {\bibinfo
  {title} {Resonant microwave-mediated interactions between distant electron
  spins},\ }\href {https://doi.org/10.1038/s41586-019-1867-y} {\bibfield
  {journal} {\bibinfo  {journal} {Nature}\ }\textbf {\bibinfo {volume} {577}},\
  \bibinfo {pages} {195} (\bibinfo {year} {2020})}\BibitemShut {NoStop}%
\bibitem [{\citenamefont {Harvey-Collard}\ \emph {et~al.}(2022)\citenamefont
  {Harvey-Collard}, \citenamefont {Dijkema}, \citenamefont {Zheng},
  \citenamefont {Sammak}, \citenamefont {Scappucci},\ and\ \citenamefont
  {Vandersypen}}]{harvey2022coherent}%
  \BibitemOpen
  \bibfield  {author} {\bibinfo {author} {\bibfnamefont {P.}~\bibnamefont
  {Harvey-Collard}}, \bibinfo {author} {\bibfnamefont {J.}~\bibnamefont
  {Dijkema}}, \bibinfo {author} {\bibfnamefont {G.}~\bibnamefont {Zheng}},
  \bibinfo {author} {\bibfnamefont {A.}~\bibnamefont {Sammak}}, \bibinfo
  {author} {\bibfnamefont {G.}~\bibnamefont {Scappucci}},\ and\ \bibinfo
  {author} {\bibfnamefont {L.~M.}\ \bibnamefont {Vandersypen}},\ }\bibfield
  {title} {\bibinfo {title} {Coherent spin-spin coupling mediated by virtual
  microwave photons},\ }\href {https://doi.org/10.1103/PhysRevX.12.021026}
  {\bibfield  {journal} {\bibinfo  {journal} {Physical Review X}\ }\textbf
  {\bibinfo {volume} {12}},\ \bibinfo {pages} {021026} (\bibinfo {year}
  {2022})}\BibitemShut {NoStop}%
\bibitem [{\citenamefont {Dijkema}\ \emph {et~al.}(2024)\citenamefont
  {Dijkema}, \citenamefont {Xue}, \citenamefont {Harvey-Collard}, \citenamefont
  {Rimbach-Russ}, \citenamefont {de~Snoo}, \citenamefont {Zheng}, \citenamefont
  {Sammak}, \citenamefont {Scappucci},\ and\ \citenamefont
  {Vandersypen}}]{dijkema2024cavity}%
  \BibitemOpen
  \bibfield  {author} {\bibinfo {author} {\bibfnamefont {J.}~\bibnamefont
  {Dijkema}}, \bibinfo {author} {\bibfnamefont {X.}~\bibnamefont {Xue}},
  \bibinfo {author} {\bibfnamefont {P.}~\bibnamefont {Harvey-Collard}},
  \bibinfo {author} {\bibfnamefont {M.}~\bibnamefont {Rimbach-Russ}}, \bibinfo
  {author} {\bibfnamefont {S.~L.}\ \bibnamefont {de~Snoo}}, \bibinfo {author}
  {\bibfnamefont {G.}~\bibnamefont {Zheng}}, \bibinfo {author} {\bibfnamefont
  {A.}~\bibnamefont {Sammak}}, \bibinfo {author} {\bibfnamefont
  {G.}~\bibnamefont {Scappucci}},\ and\ \bibinfo {author} {\bibfnamefont
  {L.~M.}\ \bibnamefont {Vandersypen}},\ }\bibfield  {title} {\bibinfo {title}
  {Cavity-mediated iswap oscillations between distant spins},\ }\href
  {https://doi.org/10.1038/s41567-024-02694-8} {\bibfield  {journal} {\bibinfo
  {journal} {Nature Physics}\ }\textbf {\bibinfo {volume} {21}},\ \bibinfo
  {pages} {168} (\bibinfo {year} {2024})}\BibitemShut {NoStop}%
\bibitem [{\citenamefont {D'Anjou}\ and\ \citenamefont
  {Burkard}(2019)}]{d2019optimal}%
  \BibitemOpen
  \bibfield  {author} {\bibinfo {author} {\bibfnamefont {B.}~\bibnamefont
  {D'Anjou}}\ and\ \bibinfo {author} {\bibfnamefont {G.}~\bibnamefont
  {Burkard}},\ }\bibfield  {title} {\bibinfo {title} {Optimal dispersive
  readout of a spin qubit with a microwave resonator},\ }\href
  {https://doi.org/10.1103/PhysRevB.100.245427} {\bibfield  {journal} {\bibinfo
   {journal} {Physical Review B}\ }\textbf {\bibinfo {volume} {100}},\ \bibinfo
  {pages} {245427} (\bibinfo {year} {2019})}\BibitemShut {NoStop}%
\bibitem [{\citenamefont {Petersson}\ \emph {et~al.}(2010)\citenamefont
  {Petersson}, \citenamefont {Smith}, \citenamefont {Anderson}, \citenamefont
  {Atkinson}, \citenamefont {Jones},\ and\ \citenamefont
  {Ritchie}}]{petersson2010charge}%
  \BibitemOpen
  \bibfield  {author} {\bibinfo {author} {\bibfnamefont {K.}~\bibnamefont
  {Petersson}}, \bibinfo {author} {\bibfnamefont {C.}~\bibnamefont {Smith}},
  \bibinfo {author} {\bibfnamefont {D.}~\bibnamefont {Anderson}}, \bibinfo
  {author} {\bibfnamefont {P.}~\bibnamefont {Atkinson}}, \bibinfo {author}
  {\bibfnamefont {G.}~\bibnamefont {Jones}},\ and\ \bibinfo {author}
  {\bibfnamefont {D.}~\bibnamefont {Ritchie}},\ }\bibfield  {title} {\bibinfo
  {title} {Charge and spin state readout of a double quantum dot coupled to a
  resonator},\ }\href {https://doi.org/10.1021/nl100663w} {\bibfield  {journal}
  {\bibinfo  {journal} {Nano {L}etters}\ }\textbf {\bibinfo {volume} {10}},\
  \bibinfo {pages} {2789} (\bibinfo {year} {2010})}\BibitemShut {NoStop}%
\bibitem [{\citenamefont {Mielke}\ \emph
  {et~al.}(2021{\natexlab{a}})\citenamefont {Mielke}, \citenamefont {Petta},\
  and\ \citenamefont {Burkard}}]{mielke2021nuclear}%
  \BibitemOpen
  \bibfield  {author} {\bibinfo {author} {\bibfnamefont {J.}~\bibnamefont
  {Mielke}}, \bibinfo {author} {\bibfnamefont {J.~R.}\ \bibnamefont {Petta}},\
  and\ \bibinfo {author} {\bibfnamefont {G.}~\bibnamefont {Burkard}},\
  }\bibfield  {title} {\bibinfo {title} {Nuclear spin readout in a
  cavity-coupled hybrid quantum dot-donor system},\ }\href
  {https://doi.org/10.1103/PRXQuantum.2.020347} {\bibfield  {journal} {\bibinfo
   {journal} {PRX Quantum}\ }\textbf {\bibinfo {volume} {2}},\ \bibinfo {pages}
  {020347} (\bibinfo {year} {2021}{\natexlab{a}})}\BibitemShut {NoStop}%
\bibitem [{\citenamefont {House}\ \emph {et~al.}(2015)\citenamefont {House},
  \citenamefont {Kobayashi}, \citenamefont {Weber}, \citenamefont {Hile},
  \citenamefont {Watson}, \citenamefont {Van Der~Heijden}, \citenamefont
  {Rogge},\ and\ \citenamefont {Simmons}}]{house2015radio}%
  \BibitemOpen
  \bibfield  {author} {\bibinfo {author} {\bibfnamefont {M.}~\bibnamefont
  {House}}, \bibinfo {author} {\bibfnamefont {T.}~\bibnamefont {Kobayashi}},
  \bibinfo {author} {\bibfnamefont {B.}~\bibnamefont {Weber}}, \bibinfo
  {author} {\bibfnamefont {S.}~\bibnamefont {Hile}}, \bibinfo {author}
  {\bibfnamefont {T.}~\bibnamefont {Watson}}, \bibinfo {author} {\bibfnamefont
  {J.}~\bibnamefont {Van Der~Heijden}}, \bibinfo {author} {\bibfnamefont
  {S.}~\bibnamefont {Rogge}},\ and\ \bibinfo {author} {\bibfnamefont
  {M.}~\bibnamefont {Simmons}},\ }\bibfield  {title} {\bibinfo {title} {Radio
  frequency measurements of tunnel couplings and singlet--triplet spin states
  in {S}i:{P} quantum dots},\ }\href {https://doi.org/10.1038/ncomms9848}
  {\bibfield  {journal} {\bibinfo  {journal} {Nature Communications}\ }\textbf
  {\bibinfo {volume} {6}},\ \bibinfo {pages} {8848} (\bibinfo {year}
  {2015})}\BibitemShut {NoStop}%
\bibitem [{\citenamefont {Colless}\ \emph {et~al.}(2013)\citenamefont
  {Colless}, \citenamefont {Mahoney}, \citenamefont {Hornibrook}, \citenamefont
  {Doherty}, \citenamefont {Lu}, \citenamefont {Gossard},\ and\ \citenamefont
  {Reilly}}]{colless2013dispersive}%
  \BibitemOpen
  \bibfield  {author} {\bibinfo {author} {\bibfnamefont {J.}~\bibnamefont
  {Colless}}, \bibinfo {author} {\bibfnamefont {A.}~\bibnamefont {Mahoney}},
  \bibinfo {author} {\bibfnamefont {J.}~\bibnamefont {Hornibrook}}, \bibinfo
  {author} {\bibfnamefont {A.}~\bibnamefont {Doherty}}, \bibinfo {author}
  {\bibfnamefont {H.}~\bibnamefont {Lu}}, \bibinfo {author} {\bibfnamefont
  {A.}~\bibnamefont {Gossard}},\ and\ \bibinfo {author} {\bibfnamefont
  {D.}~\bibnamefont {Reilly}},\ }\bibfield  {title} {\bibinfo {title}
  {Dispersive readout of a few-electron double quantum dot with fast rf gate
  sensors},\ }\href {https://doi.org/10.1103/PhysRevLett.110.046805} {\bibfield
   {journal} {\bibinfo  {journal} {Physical review letters}\ }\textbf {\bibinfo
  {volume} {110}},\ \bibinfo {pages} {046805} (\bibinfo {year}
  {2013})}\BibitemShut {NoStop}%
\bibitem [{\citenamefont {Yu}\ \emph {et~al.}(2023)\citenamefont {Yu},
  \citenamefont {Zihlmann}, \citenamefont {Abadillo-Uriel}, \citenamefont
  {Michal}, \citenamefont {Rambal}, \citenamefont {Niebojewski}, \citenamefont
  {Bedecarrats}, \citenamefont {Vinet}, \citenamefont {Dumur}, \citenamefont
  {Filippone} \emph {et~al.}}]{GE_Floppingmode_Exp_yu2023strong}%
  \BibitemOpen
  \bibfield  {author} {\bibinfo {author} {\bibfnamefont {C.~X.}\ \bibnamefont
  {Yu}}, \bibinfo {author} {\bibfnamefont {S.}~\bibnamefont {Zihlmann}},
  \bibinfo {author} {\bibfnamefont {J.~C.}\ \bibnamefont {Abadillo-Uriel}},
  \bibinfo {author} {\bibfnamefont {V.~P.}\ \bibnamefont {Michal}}, \bibinfo
  {author} {\bibfnamefont {N.}~\bibnamefont {Rambal}}, \bibinfo {author}
  {\bibfnamefont {H.}~\bibnamefont {Niebojewski}}, \bibinfo {author}
  {\bibfnamefont {T.}~\bibnamefont {Bedecarrats}}, \bibinfo {author}
  {\bibfnamefont {M.}~\bibnamefont {Vinet}}, \bibinfo {author} {\bibfnamefont
  {{\'E}.}~\bibnamefont {Dumur}}, \bibinfo {author} {\bibfnamefont
  {M.}~\bibnamefont {Filippone}}, \emph {et~al.},\ }\bibfield  {title}
  {\bibinfo {title} {Strong coupling between a photon and a hole spin in
  silicon},\ }\href {https://doi.org/10.1038/s41565-023-01332-3} {\bibfield
  {journal} {\bibinfo  {journal} {Nature {N}anotechnology}\ }\textbf {\bibinfo
  {volume} {18}},\ \bibinfo {pages} {741} (\bibinfo {year} {2023})}\BibitemShut
  {NoStop}%
\bibitem [{\citenamefont {Delva}\ \emph {et~al.}(2024)\citenamefont {Delva},
  \citenamefont {Mielke}, \citenamefont {Burkard},\ and\ \citenamefont
  {Petta}}]{Petta_Jonas_entagledFloppingmodesPhysRevB.110.035304}%
  \BibitemOpen
  \bibfield  {author} {\bibinfo {author} {\bibfnamefont {R.~L.}\ \bibnamefont
  {Delva}}, \bibinfo {author} {\bibfnamefont {J.}~\bibnamefont {Mielke}},
  \bibinfo {author} {\bibfnamefont {G.}~\bibnamefont {Burkard}},\ and\ \bibinfo
  {author} {\bibfnamefont {J.~R.}\ \bibnamefont {Petta}},\ }\bibfield  {title}
  {\bibinfo {title} {Measurement-based entanglement of semiconductor spin
  qubits},\ }\href {https://doi.org/10.1103/PhysRevB.110.035304} {\bibfield
  {journal} {\bibinfo  {journal} {Phys. Rev. B}\ }\textbf {\bibinfo {volume}
  {110}},\ \bibinfo {pages} {035304} (\bibinfo {year} {2024})}\BibitemShut
  {NoStop}%
\bibitem [{\citenamefont {Mutter}\ and\ \citenamefont
  {Burkard}(2021)}]{FloppingmodeinGE_Mutter_-PhysRevResearch.3.013194}%
  \BibitemOpen
  \bibfield  {author} {\bibinfo {author} {\bibfnamefont {P.~M.}\ \bibnamefont
  {Mutter}}\ and\ \bibinfo {author} {\bibfnamefont {G.}~\bibnamefont
  {Burkard}},\ }\bibfield  {title} {\bibinfo {title} {Natural heavy-hole
  flopping mode qubit in germanium},\ }\href
  {https://doi.org/10.1103/PhysRevResearch.3.013194} {\bibfield  {journal}
  {\bibinfo  {journal} {Phys. Rev. Res.}\ }\textbf {\bibinfo {volume} {3}},\
  \bibinfo {pages} {013194} (\bibinfo {year} {2021})}\BibitemShut {NoStop}%
\bibitem [{\citenamefont {Hu}\ \emph {et~al.}(2023)\citenamefont {Hu},
  \citenamefont {Ma}, \citenamefont {Ni}, \citenamefont {Zhou}, \citenamefont
  {Chu}, \citenamefont {Liao}, \citenamefont {Kong}, \citenamefont {Cao},
  \citenamefont {Wang}, \citenamefont {Li},\ and\ \citenamefont
  {Guo}}]{FloppingExpChina10.1063/5.0137259}%
  \BibitemOpen
  \bibfield  {author} {\bibinfo {author} {\bibfnamefont {R.-Z.}\ \bibnamefont
  {Hu}}, \bibinfo {author} {\bibfnamefont {R.-L.}\ \bibnamefont {Ma}}, \bibinfo
  {author} {\bibfnamefont {M.}~\bibnamefont {Ni}}, \bibinfo {author}
  {\bibfnamefont {Y.}~\bibnamefont {Zhou}}, \bibinfo {author} {\bibfnamefont
  {N.}~\bibnamefont {Chu}}, \bibinfo {author} {\bibfnamefont {W.-Z.}\
  \bibnamefont {Liao}}, \bibinfo {author} {\bibfnamefont {Z.-Z.}\ \bibnamefont
  {Kong}}, \bibinfo {author} {\bibfnamefont {G.}~\bibnamefont {Cao}}, \bibinfo
  {author} {\bibfnamefont {G.-L.}\ \bibnamefont {Wang}}, \bibinfo {author}
  {\bibfnamefont {H.-O.}\ \bibnamefont {Li}},\ and\ \bibinfo {author}
  {\bibfnamefont {G.-P.}\ \bibnamefont {Guo}},\ }\bibfield  {title} {\bibinfo
  {title} {Flopping-mode spin qubit in a {S}i-{MOS} quantum dot},\ }\href
  {https://doi.org/10.1063/5.0137259} {\bibfield  {journal} {\bibinfo
  {journal} {Applied Physics Letters}\ }\textbf {\bibinfo {volume} {122}},\
  \bibinfo {pages} {134002} (\bibinfo {year} {2023})}\BibitemShut {NoStop}%
\bibitem [{\citenamefont {Ginzel}\ and\ \citenamefont
  {Burkard}(2022)}]{PhysRevResearch.4.033048}%
  \BibitemOpen
  \bibfield  {author} {\bibinfo {author} {\bibfnamefont {F.}~\bibnamefont
  {Ginzel}}\ and\ \bibinfo {author} {\bibfnamefont {G.}~\bibnamefont
  {Burkard}},\ }\bibfield  {title} {\bibinfo {title} {Proposal for a
  cavity-induced measurement of the exchange coupling in quantum dots},\ }\href
  {https://doi.org/10.1103/PhysRevResearch.4.033048} {\bibfield  {journal}
  {\bibinfo  {journal} {Phys. Rev. Res.}\ }\textbf {\bibinfo {volume} {4}},\
  \bibinfo {pages} {033048} (\bibinfo {year} {2022})}\BibitemShut {NoStop}%
\bibitem [{\citenamefont {Blais}\ \emph {et~al.}(2021)\citenamefont {Blais},
  \citenamefont {Grimsmo}, \citenamefont {Girvin},\ and\ \citenamefont
  {Wallraff}}]{blais2021circuit}%
  \BibitemOpen
  \bibfield  {author} {\bibinfo {author} {\bibfnamefont {A.}~\bibnamefont
  {Blais}}, \bibinfo {author} {\bibfnamefont {A.~L.}\ \bibnamefont {Grimsmo}},
  \bibinfo {author} {\bibfnamefont {S.~M.}\ \bibnamefont {Girvin}},\ and\
  \bibinfo {author} {\bibfnamefont {A.}~\bibnamefont {Wallraff}},\ }\bibfield
  {title} {\bibinfo {title} {Circuit quantum electrodynamics},\ }\href
  {https://doi.org/10.1103/RevModPhys.93.025005} {\bibfield  {journal}
  {\bibinfo  {journal} {Reviews of Modern Physics}\ }\textbf {\bibinfo {volume}
  {93}},\ \bibinfo {pages} {025005} (\bibinfo {year} {2021})}\BibitemShut
  {NoStop}%
\bibitem [{\citenamefont {Didier}\ \emph {et~al.}(2015)\citenamefont {Didier},
  \citenamefont {Bourassa},\ and\ \citenamefont {Blais}}]{didier2015fast}%
  \BibitemOpen
  \bibfield  {author} {\bibinfo {author} {\bibfnamefont {N.}~\bibnamefont
  {Didier}}, \bibinfo {author} {\bibfnamefont {J.}~\bibnamefont {Bourassa}},\
  and\ \bibinfo {author} {\bibfnamefont {A.}~\bibnamefont {Blais}},\ }\bibfield
   {title} {\bibinfo {title} {Fast quantum nondemolition readout by parametric
  modulation of longitudinal qubit-oscillator interaction},\ }\href
  {https://doi.org/10.1103/PhysRevLett.115.203601} {\bibfield  {journal}
  {\bibinfo  {journal} {Physical review letters}\ }\textbf {\bibinfo {volume}
  {115}},\ \bibinfo {pages} {203601} (\bibinfo {year} {2015})}\BibitemShut
  {NoStop}%
\bibitem [{\citenamefont {Jin}\ \emph {et~al.}(2012)\citenamefont {Jin},
  \citenamefont {Marthaler}, \citenamefont {Shnirman},\ and\ \citenamefont
  {Sch{\"o}n}}]{jin2012strong}%
  \BibitemOpen
  \bibfield  {author} {\bibinfo {author} {\bibfnamefont {P.-Q.}\ \bibnamefont
  {Jin}}, \bibinfo {author} {\bibfnamefont {M.}~\bibnamefont {Marthaler}},
  \bibinfo {author} {\bibfnamefont {A.}~\bibnamefont {Shnirman}},\ and\
  \bibinfo {author} {\bibfnamefont {G.}~\bibnamefont {Sch{\"o}n}},\ }\bibfield
  {title} {\bibinfo {title} {Strong coupling of spin qubits to a transmission
  line resonator},\ }\href {https://doi.org/10.1103/PhysRevLett.108.190506}
  {\bibfield  {journal} {\bibinfo  {journal} {Physical review letters}\
  }\textbf {\bibinfo {volume} {108}},\ \bibinfo {pages} {190506} (\bibinfo
  {year} {2012})}\BibitemShut {NoStop}%
\bibitem [{\citenamefont {Richer}\ \emph {et~al.}(2017)\citenamefont {Richer},
  \citenamefont {Maleeva}, \citenamefont {Skacel}, \citenamefont {Pop},\ and\
  \citenamefont {DiVincenzo}}]{transmonricher2017inductively}%
  \BibitemOpen
  \bibfield  {author} {\bibinfo {author} {\bibfnamefont {S.}~\bibnamefont
  {Richer}}, \bibinfo {author} {\bibfnamefont {N.}~\bibnamefont {Maleeva}},
  \bibinfo {author} {\bibfnamefont {S.~T.}\ \bibnamefont {Skacel}}, \bibinfo
  {author} {\bibfnamefont {I.~M.}\ \bibnamefont {Pop}},\ and\ \bibinfo {author}
  {\bibfnamefont {D.}~\bibnamefont {DiVincenzo}},\ }\bibfield  {title}
  {\bibinfo {title} {Inductively shunted transmon qubit with tunable transverse
  and longitudinal coupling},\ }\href
  {https://doi.org/10.1103/PhysRevB.96.174520} {\bibfield  {journal} {\bibinfo
  {journal} {Physical Review B}\ }\textbf {\bibinfo {volume} {96}},\ \bibinfo
  {pages} {174520} (\bibinfo {year} {2017})}\BibitemShut {NoStop}%
\bibitem [{\citenamefont {Ruskov}\ and\ \citenamefont
  {Tahan}(2019)}]{ruskov2019quantum}%
  \BibitemOpen
  \bibfield  {author} {\bibinfo {author} {\bibfnamefont {R.}~\bibnamefont
  {Ruskov}}\ and\ \bibinfo {author} {\bibfnamefont {C.}~\bibnamefont {Tahan}},\
  }\bibfield  {title} {\bibinfo {title} {Quantum-limited measurement of spin
  qubits via curvature couplings to a cavity},\ }\href
  {https://doi.org/10.1103/PhysRevB.99.245306} {\bibfield  {journal} {\bibinfo
  {journal} {Physical Review B}\ }\textbf {\bibinfo {volume} {99}},\ \bibinfo
  {pages} {245306} (\bibinfo {year} {2019})}\BibitemShut {NoStop}%
\bibitem [{\citenamefont {Ruskov}\ and\ \citenamefont
  {Tahan}(2021)}]{ruskov2021modulated}%
  \BibitemOpen
  \bibfield  {author} {\bibinfo {author} {\bibfnamefont {R.}~\bibnamefont
  {Ruskov}}\ and\ \bibinfo {author} {\bibfnamefont {C.}~\bibnamefont {Tahan}},\
  }\bibfield  {title} {\bibinfo {title} {Modulated longitudinal gates on
  encoded spin qubits via curvature couplings to a superconducting cavity},\
  }\href {https://doi.org/10.1103/PhysRevB.103.035301} {\bibfield  {journal}
  {\bibinfo  {journal} {Physical Review B}\ }\textbf {\bibinfo {volume}
  {103}},\ \bibinfo {pages} {035301} (\bibinfo {year} {2021})}\BibitemShut
  {NoStop}%
\bibitem [{\citenamefont {Estakhri}\ \emph {et~al.}(2024)\citenamefont
  {Estakhri}, \citenamefont {Warren}, \citenamefont {Economou},\ and\
  \citenamefont {Barnes}}]{TDQresonatorsehrahnlich_PhysRevResearch.6.043029}%
  \BibitemOpen
  \bibfield  {author} {\bibinfo {author} {\bibfnamefont {N.~M.}\ \bibnamefont
  {Estakhri}}, \bibinfo {author} {\bibfnamefont {A.}~\bibnamefont {Warren}},
  \bibinfo {author} {\bibfnamefont {S.~E.}\ \bibnamefont {Economou}},\ and\
  \bibinfo {author} {\bibfnamefont {E.}~\bibnamefont {Barnes}},\ }\bibfield
  {title} {\bibinfo {title} {Long-distance photon-mediated and short-distance
  entangling gates in three-qubit quantum dot spin systems},\ }\href
  {https://doi.org/10.1103/PhysRevResearch.6.043029} {\bibfield  {journal}
  {\bibinfo  {journal} {Phys. Rev. Res.}\ }\textbf {\bibinfo {volume} {6}},\
  \bibinfo {pages} {043029} (\bibinfo {year} {2024})}\BibitemShut {NoStop}%
\bibitem [{\citenamefont {Gardiner}\ and\ \citenamefont
  {Collett}(1985)}]{Gardiner1985}%
  \BibitemOpen
  \bibfield  {author} {\bibinfo {author} {\bibfnamefont {C.~W.}\ \bibnamefont
  {Gardiner}}\ and\ \bibinfo {author} {\bibfnamefont {M.~J.}\ \bibnamefont
  {Collett}},\ }\bibfield  {title} {\bibinfo {title} {Input and output in
  damped quantum systems: Quantum stochastic differential equations and the
  master equation},\ }\href {https://doi.org/10.1103/PhysRevA.31.3761}
  {\bibfield  {journal} {\bibinfo  {journal} {Phys. Rev. A}\ }\textbf {\bibinfo
  {volume} {31}},\ \bibinfo {pages} {3761} (\bibinfo {year}
  {1985})}\BibitemShut {NoStop}%
\bibitem [{\citenamefont {Hsieh}\ \emph {et~al.}(2012)\citenamefont {Hsieh},
  \citenamefont {Shim}, \citenamefont {Korkusinski},\ and\ \citenamefont
  {Hawrylak}}]{hsieh2012physics}%
  \BibitemOpen
  \bibfield  {author} {\bibinfo {author} {\bibfnamefont {C.-Y.}\ \bibnamefont
  {Hsieh}}, \bibinfo {author} {\bibfnamefont {Y.-P.}\ \bibnamefont {Shim}},
  \bibinfo {author} {\bibfnamefont {M.}~\bibnamefont {Korkusinski}},\ and\
  \bibinfo {author} {\bibfnamefont {P.}~\bibnamefont {Hawrylak}},\ }\bibfield
  {title} {\bibinfo {title} {Physics of lateral triple quantum-dot molecules
  with controlled electron numbers},\ }\href
  {https://doi.org/10.1088/0034-4885/75/11/114501} {\bibfield  {journal}
  {\bibinfo  {journal} {Reports on Progress in Physics}\ }\textbf {\bibinfo
  {volume} {75}},\ \bibinfo {pages} {114501} (\bibinfo {year}
  {2012})}\BibitemShut {NoStop}%
\bibitem [{\citenamefont {Ginzel}\ and\ \citenamefont
  {Burkard}(2023)}]{GinzelreadoutPhysRevB.108.125437}%
  \BibitemOpen
  \bibfield  {author} {\bibinfo {author} {\bibfnamefont {F.}~\bibnamefont
  {Ginzel}}\ and\ \bibinfo {author} {\bibfnamefont {G.}~\bibnamefont
  {Burkard}},\ }\bibfield  {title} {\bibinfo {title} {Simultaneous transient
  dispersive readout of multiple spin qubits},\ }\href
  {https://doi.org/10.1103/PhysRevB.108.125437} {\bibfield  {journal} {\bibinfo
   {journal} {Phys. Rev. B}\ }\textbf {\bibinfo {volume} {108}},\ \bibinfo
  {pages} {125437} (\bibinfo {year} {2023})}\BibitemShut {NoStop}%
\bibitem [{\citenamefont {Mielke}\ \emph
  {et~al.}(2021{\natexlab{b}})\citenamefont {Mielke}, \citenamefont {Petta},\
  and\ \citenamefont {Burkard}}]{JonasMielkePRXQuantum.2.020347}%
  \BibitemOpen
  \bibfield  {author} {\bibinfo {author} {\bibfnamefont {J.}~\bibnamefont
  {Mielke}}, \bibinfo {author} {\bibfnamefont {J.~R.}\ \bibnamefont {Petta}},\
  and\ \bibinfo {author} {\bibfnamefont {G.}~\bibnamefont {Burkard}},\
  }\bibfield  {title} {\bibinfo {title} {Nuclear spin readout in a
  cavity-coupled hybrid quantum dot-donor system},\ }\href
  {https://doi.org/10.1103/PRXQuantum.2.020347} {\bibfield  {journal} {\bibinfo
   {journal} {PRX Quantum}\ }\textbf {\bibinfo {volume} {2}},\ \bibinfo {pages}
  {020347} (\bibinfo {year} {2021}{\natexlab{b}})}\BibitemShut {NoStop}%
\end{thebibliography}%

\end{document}